\documentclass[10pt, draftclsnofoot, twocolumn]{IEEEtran}
\pdfoutput=1
\usepackage{amssymb}
\usepackage{algorithm}
\usepackage{algpseudocode}
\usepackage{algorithmicx}
\usepackage{amsfonts}
\usepackage{amsmath}
\usepackage{amssymb}
\usepackage{amsthm}
\usepackage{array}
\usepackage{bbding}
\usepackage{bigints}
\usepackage{booktabs}
\usepackage{cite}
\usepackage{geometry}
\usepackage{color}
\usepackage{diagbox}
\usepackage{epsfig,latexsym}
\usepackage{epstopdf}
\usepackage{graphicx}
\usepackage{fancyhdr}
\usepackage{float}
\usepackage{indentfirst}
\usepackage{lastpage}
\usepackage{makecell}
\usepackage{mathtools}
\usepackage{multirow}
\usepackage{pifont}
\usepackage{psfrag}
\usepackage{setspace}
\usepackage{stfloats}
\usepackage[colorlinks,linkcolor=blue]{hyperref}
\hypersetup{hidelinks,
	colorlinks=true,
	allcolors=black,
	pdfstartview=Fit,
	breaklinks=true}
\usepackage{cleveref}
\renewcommand{\baselinestretch}{1.0}
\usepackage{empheq}
\usepackage{enumerate}
\usepackage{times}
\usepackage{subfigure}
\newcolumntype{C}{>{\centering\arraybackslash$}p{\linewidth}<{$}}

\newtheorem{theorem}{Theorem}

\newtheorem{lemma}{Lemma}

\newtheorem{corollary}{Corollary}

\DeclareMathOperator{\diag}{diag}
\DeclareMathOperator{\tr}{tr}

\newcommand{\RNum}[1]{\uppercase\expandafter{\romannumeral #1\relax}}

\newcounter{eqncnt}

\newcounter{eqnback}

\begin{document}

\title{RIS-Assisted Cell-Free Massive MIMO Relying on Reflection Pattern Modulation}

\author{{Zeping Sui}, {\em Member,~IEEE}, {Hien Quoc Ngo}, {\em Senior Member,~IEEE}, {Trinh Van Chien}, {\em Member,~IEEE}, {Michail Matthaiou}, {\em Fellow,~IEEE}, and {Lajos Hanzo}, {\em Life Fellow,~IEEE}
\vspace{-1em}
\thanks{This work is a contribution by Project REASON, a UK Government funded project under the Future Open Networks Research Challenge (FONRC) sponsored by the Department of Science Innovation and Technology (DSIT). It was also supported by the U.K. Engineering and Physical Sciences Research Council (EPSRC) (grant No. EP/X04047X/1). The work of M. Matthaiou was supported by the European Research Council (ERC) under the European Union’s Horizon 2020 research and innovation programme (grant agreement No. 101001331).}
\thanks{L. Hanzo would like to acknowledge the financial support of the Engineering and Physical Sciences Research Council projects EP/W016605/1, EP/X01228X/1, EP/Y026721/1, EP/W032635/1 and EP/X04047X/1 as well as of the European Research Council's Advanced Fellow Grant QuantCom (Grant No. 789028)}
\thanks{Zeping Sui, Hien Quoc Ngo and Michail Matthaiou are with the Centre for Wireless Innovation (CWI), Queen's University Belfast, Belfast BT3 9DT, U.K. (e-mail: \{z.sui, hien.ngo, m.matthaiou\}@qub.ac.uk).}
\thanks{Trinh Van Chien is with the School of Information and Communication Technology, Hanoi University of Science and Technology, Hanoi, 10000, Vietnam (email: chientv@soict.hust.edu.vn).}
\thanks{Lajos Hanzo is with the Department of Electronics and Computer Science, University of Southampton, Southampton SO17 1BJ, U.K. (e-mail: lh@ecs.soton.ac.uk).}%
}
\maketitle

\begin{abstract}
We propose reflection pattern modulation-aided reconfigurable intelligent surface (RPM-RIS)-assisted cell-free massive multiple-input-multiple-output (CF-mMIMO) schemes for green uplink transmission. In our RPM-RIS-assisted CF-mMIMO system, extra information is conveyed by the indices of the active RIS blocks, exploiting the joint benefits of both RIS-assisted CF-mMIMO transmission and RPM. Since only part of the RIS blocks are active, our proposed architecture strikes a flexible energy \emph{vs.} spectral efficiency (SE) trade-off. We commence with introducing the system model by considering spatially correlated channels. Moreover, we conceive a channel estimation scheme subject to the linear minimum mean-square error (MMSE) constraint, yielding sufficient information for the subsequent signal processing steps. Then, upon exploiting a so-called large-scale fading decoding (LSFD) scheme, the uplink signal-to-interference-and-noise ratio (SINR) is derived based on the RIS ON/OFF statistics, where both maximum ratio (MR) and local minimum mean-square error (L-MMSE) combiners are considered. By invoking the MR combiner, the closed-form expression of the uplink SE is formulated based only on the channel statistics. Furthermore, we derive the total energy efficiency (EE) of our proposed RPM-RIS-assisted CF-mMIMO system. Additionally, we propose a chaotic sequence-based adaptive particle swarm optimization (CSA-PSO) algorithm to maximize the total EE by designing the RIS phase shifts. Specifically, the initial particle diversity is promoted by invoking chaotic sequences, and an adaptive time-varying inertia weight is developed to improve its particle search performance. Furthermore, the particle mutation and reset steps are appropriately selected to enable the algorithm to escape from local optima. Finally, our simulation results demonstrate that the proposed RPM-RIS-assisted CF-mMIMO architecture strikes an attractive SE \emph{vs.} EE trade-off, while the CSA-PSO algorithm is capable of attaining a significant EE performance gain compared to conventional solutions.
\end{abstract}
\begin{IEEEkeywords}
Cell-free massive MIMO, energy efficiency, iterative optimization, reconfigurable intelligent surfaces, reflection pattern modulation, spectral efficiency.
\end{IEEEkeywords}
\IEEEpeerreviewmaketitle

\section{Introduction}\label{Section 1}
Cell-free massive multiple-input-multiple-output (CF-mMIMO) schemes constitute promising candidates for next-generation wireless communications \cite{7827017,8097026,zhang2020prospective}. Specifically, a large number of access points (APs) are randomly distributed in a given area to support many user equipment (UE), which are connected to one or several central processing units (CPUs) via backhaul links \cite{bjornson2019making,10522673}. Compared to conventional cellular mMIMO systems, where the cell-edge UEs suffer from poor quality-of-service (QoS) and severe inter-cell interference \cite{6482234,9354156}, a CF-mMIMO system experiences reduced interference and excellent load-balancing, yielding improved spectral efficiency (SE), lower transmission latency and increased network capacity \cite{7827017,8097026,bjornson2019making}. However, under propagation scenarios with poor scattering and/or high transmission path loss, the SE and EE of CF-mMIMO will be compromised \cite{10197459}. As a remedy, reconfigurable intelligent surfaces (RISs) have been widely integrated with various communication scenarios \cite{9614196,li2022reconfigurable}. Specifically, a RIS includes a large number of passive elements coordinated by a micro-controller, while the reflected signals can be shaped in terms of their phase and magnitude at the electromagnetic level, resulting in passive beamforming \cite{8910627}. Since no active beamformers and sophisticated signal processing techniques are needed, this results in decreased overall energy consumption \cite{huang2019reconfigurable}.

More recently, RIS-assisted CF-mMIMO systems have been widely investigated in the existing literature \cite{van2021reconfigurable,10167480,10197459,10326460,10129196,shi2022wireless}. Specifically, in \cite{van2021reconfigurable}, both the uplink and downlink SEs of RIS-assisted CF-mMIMO systems was studied based on maximum ratio (MR) processing and the estimated cascaded two-hop channels were derived, where single-antenna APs were considered. Shi \emph{et al.} \cite{10167480} considered the electromagnetic interference (EMI) in RIS-assisted CF-mMIMO systems, and the uplink SE was also derived. The uplink SE and EE of a RIS-assisted CF-mMIMO system having realistic hardware impairments were investigated in \cite{10197459}, while a joint max-min SE optimization paradigm of both large-scale fading decoding (LSFD) and power control was also proposed. A two-timescale transmission protocol-based RIS-assisted CF-mMIMO system was proposed in \cite{10326460}, where the RIS-based passive beamforming was designed relying on the channel statistics and the MR combiner was formulated based on the instantaneous channels. In \cite{10129196}, Yao \emph{et al.} investigated a robust beamforming scheme of a RIS-assisted CF-mMIMO system by incorporating the realistic channel state information (CSI) uncertainties, where a limited-rate backhaul was considered. {In \cite{shi2022wireless}, a comprehensive literature survey of wireless energy transfer in RIS-aided CF-mMIMO systems was carried out. Both the downlink and uplink SE of RIS-aided CF-mMIMO systems with channel aging were analyzed in \cite{10468556}. Diluka \emph{et al.} derived tight bounds of the achievable rate and outage probability in closed-form for RIS-assisted cell-free systems in \cite{9685730}. Later in \cite{9473521}, the authors investigated RIS-aided CF-mMIMO systems relying on single-antenna APs. Moreover, a closed-form expression of the achievable rate was derived. In \cite{9459505}, upon considering a wideband RIS-aided cell-free network, a joint precoding scheme was conceived for maximizing the network capacity. On the other hand, the maximum EE optimization of RIS-aided cell-free networks was disseminated in \cite{10290997,10175060,9363171,10035459}. Explicitly, limited backhaul capacity was considered in \cite{9363171}, while hybrid RIS-based systems were studied in \cite{10035459}.} However, the above-mentioned studies consider RISs only as passive reflectors to reshape the propagation channels. It is known from \cite{9499059} that extra information can also be transmitted by a RIS, leading to an improved communication data rate, which is close to the RIS channel capacity in theory.

As a parallel development, reflection modulation (RM) has been proposed to achieve additional information transmission \cite{8941126,9133134}. In \cite{8941126}, a passive beamforming and information transfer diagram was proposed. In more details, extra bits can be mapped onto the ON/OFF states of the RIS elements, while passive beamforming can be achieved by appropriately adjusting the active phase shifts. However, since the number of active RIS elements changes among the transmitted symbols, the outage probability performance of passive beamforming and information transfer is compromised \cite{9133134}. To circument this problem, Lin \emph{et al.} \cite{9217944} proposed reflection pattern modulation (RPM), where the RIS reflection pattern is fixed during each symbol transmission, thereby enhancing the outage probability compared to that of the conventional passive beamforming and information transfer scheme. In this context, a received signal power \emph{vs.} achievable rate trade-off can also be formulated. {Then the RPM system was extended to quadrature reflection modulation in \cite{9516949}, where all RIS elements are activated. In \cite{10224829}, extra information bits were mapped onto the superimposed RIS phase shifts. Upon dividing the RIS into active and passive blocks, a novel hybrid reflection modulation scheme was proposed in \cite{9965423}.} However, the above-mentioned design philosophy has not been harnessed for RIS-assisted CF-mMIMO systems. Moreover, by scanning the related literature, we can conjecture that the performance of RIS-assisted CF-mMIMO systems can be further improved by leveraging RPM techniques. Against this backdrop, we activate part of the RIS elements for supporting the uplink transmission of a RIS-assisted CF-mMIMO system over Rician spatially correlated channels. Explicitly, by intrinsically amalgamating RPM and RIS-assisted CF-mMIMO, we propose the RPM-RIS-assisted CF-mMIMO philosophy for green communications.

The contributions of this paper are boldly contrasted to the existing literature in Table \ref{table1}, which are itemized below:
\begin{table*}[t]
\footnotesize
\centering
\caption{Contrasting the contributions to state-of-the-art}
\label{table1}
\begin{tabular}{l|c|c|c|c|c|c|c}
\hline
Contributions & \textbf{This paper} & \cite{10197459} & \cite{van2021reconfigurable} & \cite{10167480} & \cite{10326460} & \cite{10129196} & \cite{8941126,9133134,9217944}  \\
\hline
\hline
RIS-assisted CF-mMIMO & \checkmark & \checkmark & \checkmark  & \checkmark & \checkmark & \checkmark &  \\  
\hline
RPM & \checkmark &  &  &   & &  & \checkmark  \\  
\hline
Spatially correlated Rician channels & \checkmark &  &  &  \checkmark & & & \\  
\hline
LSFD cooperation & \checkmark & \checkmark &   &   \checkmark & & & \\  
\hline 
SE in closed-form of RPM-RIS-assisted CF-mMIMO & \checkmark &    &   &  &   & & \\  
\hline
EE analysis & \checkmark & \checkmark &  &   & \checkmark & & \\ 
\hline
Max EE-based RIS phase shift design & \checkmark &  &  &   & \checkmark & & \\ 
\hline
PSO-based EE optimization & \checkmark &  &  &   &  & & \\ 
\hline
Complexity analysis of PSO-like algorithms & \checkmark &  &   &  & & &   \\ 
\hline
\end{tabular}
\vspace{-3em}
\end{table*}
{\begin{itemize}
\item We propose an RPM-RIS-assisted CF-mMIMO scheme for supporting green uplink transmission over spatially correlated channels, where information is conveyed both by the classic amplitude-phase modulated (APM) symbols and the indices of the active RIS elements. The RIS-reflected channel is modeled by the correlated Rician fading distribution. Then, the uplink linear minimum mean-square error (MMSE) cascaded two-hop channel estimation algorithm is presented, where realistic pilot contamination is considered and {yields sufficient information for sequential data processing.} Moreover, the LSFD cooperation technique is exploited to formulate the uplink input-output relationship, where both the MR and local MMSE (L-MMSE) combiners are invoked.
\item Based on the MR combiner and the RIS ON/OFF statistics, a closed-form expression is derived for the uplink SE for finite numbers of UEs and APs. Explicitly, the effects of pilot contamination, channel estimation error, and RIS phase shifts are considered. Furthermore, the amount of information contained in the received signals is derived by invoking entropy theory. Upon utilizing the results derived, the total EE of our proposed RPM-RIS-assisted CF-mMIMO system is investigated. The simulation results demonstrate that the proposed RPM-RIS-assisted CF-mMIMO system is capable of approaching the SE of the conventional RIS-assisted CF-mMIMO. Moreover, it can also be observed from the simulations that our RPM-RIS-assisted CF-mMIMO can attain better total EE than RIS-assisted CF-mMIMO, where all RIS elements are active.
\item Aiming for maximizing the total EE under a RIS phase shift constraint and per-user EE constraint, the optimization problem of RIS phase shift design is formulated. Due to the non-convexity of the formulated optimization problem, we propose a chaotic sequence-based adaptive particle swarm optimization (CSA-PSO) algorithm to solve the RIS phase shift design problem. In more detail, we use chaotic sequences to improve the initial particle diversity. Furthermore, a novel adaptive inertia weight factor is proposed for improving its search performance, and beneficial particle mutation and reset steps are conceived to help the particles escape from local optima. It is shown that our proposed CSA-PSO significantly enhances the EE compared to its conventional PSO and random phase shift based counterparts. Combined with the SE simulation results, we conclude that the proposed RPM-RIS-assisted CF-mMIMO system studies an attractive SE \emph{vs.} EE trade-off.
\end{itemize}}

The remainder of our paper is organized as follows: The system model is described in Section \ref{Section 2}, while the uplink data transmission and SE are derived in Section \ref{Section 3}. In Section \ref{Section 4}, the uplink power consumption and EE are investigated, while the RIS phase shift design formulated for uplink EE maximization is introduced in Section \ref{Section 5}. In Section \ref{Section 6}, we present our simulation results. Finally, our conclusions are discussed in Section \ref{Section 7}.

\emph{Notation:} Henceforth, $\otimes$, $\mathbb{E}\{\cdot\}$ and $\text{tr}(\cdot)$ are the Kronecker product, expectation and trace operators. Matrices and vectors are denoted by upper- and lower-case boldface letters, respectively. A complex Gaussian distribution with mean vector $\pmb{a}$ and covariance matrix $\pmb{B}$ is denoted by $\mathcal{CN}(\pmb{a},\pmb{B})$; $(\cdot)^T$, $(\cdot)^{\ast}$, $(\cdot)^H$, $(\cdot)^{-1}$ and $\left|\cdot\right|^2$ are the transpose, conjugate, conjugate transpose, inverse and magnitude operators respectively; $B(a,b)$ represents the $(a,b)$th element of the matrix $\pmb{B}$; $\pmb{I}_N$ denotes the $N$-dimensional identity matrix; $\mathcal{U}[a,b]$ represents the uniform distribution in the interval $[a,b]$. The Euclidean norm operator and truncation argument are denoted by $\left \| \cdot \right \|$ and $\left \lfloor \cdot \right \rfloor$, respectively. Finally, $\pmb{A}=\text{diag}\{\pmb{a}\}$ denotes a diagonal matrix $\pmb{A}$ with the elements of the vector $\pmb{a}$ on the diagonal.
\section{System Model and Channel Estimation}\label{Section 2}
\begin{figure}[t]
\centering
\includegraphics[width=0.9\linewidth]{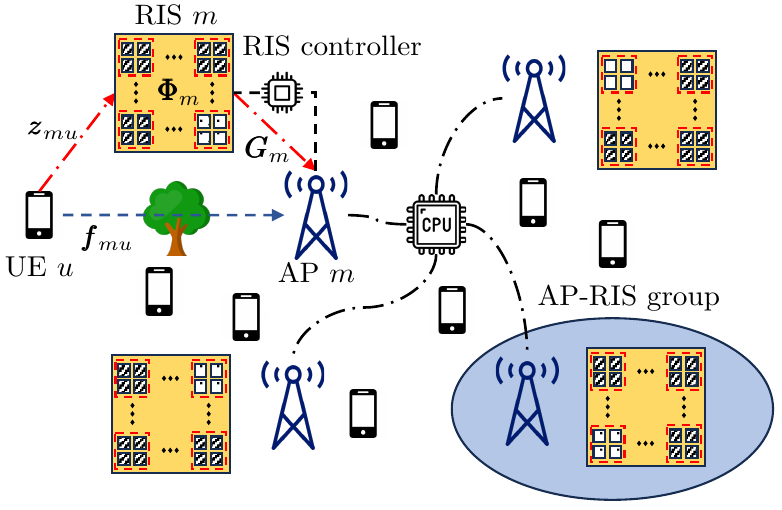}
\caption{Illustration of the RPM-RIS-assisted CF-mMIMO system employing $M$ AP-RIS groups and $U$ UEs.}
\label{Figure1}
\end{figure}

Let us consider a centralized RPM-RIS-assisted CF-mMIMO communication system that involves $U$ single-antenna UEs, $M$ APs and $M$ RISs, as shown in Fig. \ref{Figure1}. Specifically, each RIS includes $L$ elements and each AP is equipped with $J$ antennas. Moreover, each RIS is associated with an RIS controller, capable of adjusting the phase shift and reflection amplitude of all RIS elements, while receiving information symbols from UE \cite{9217944}.\footnote{{Since our proposed channel estimation method is conceived for the \emph{cascaded channels} only in terms of statistical CSI, it is straightforward to extend our architecture to generalized systems in which the number of APs and RISs are different, as shown in \cite{10167480,van2021reconfigurable}.}} We assume that the locations of all the APs and users are randomly chosen in a specific area, and the time-division duplexing (TDD) operation is adopted similarly to conventional CF-mMIMO systems \cite{7827017}, yielding channel reciprocity between the uplink and downlink transmissions. By utilizing RIS beamforming, the reflected signal of one RIS can be directed to its associated AP \cite{9110889,9779130}, i.e., each AP is served by only one RIS that is employed close to the corresponding AP. All the RISs are divided into $G$ blocks, and each block has $N=L/G$ elements, yielding overall $L_A=NK$ active elements. Now, we focus our attention on the $m$-th AP-RIS group. At the beginning of each channel coherence interval, $L_1=\left \lfloor \log_2\tbinom{G}{K} \right \rfloor$ bits are mapped onto $K$ activated RIS block index symbols, which are then transmitted to the corresponding RIS controller to activate specific RIS blocks. Briefly, there are only $K$ out of $G$ blocks that stay activated (ON-state), and the remaining $(G-K)$ are OFF-state blocks. Hence, there are a total of $C=2^{L_1}$ reflection patterns (RPs), which can be expressed as $\mathcal{I}_m=\{\mathcal{I}_{m,1},\ldots,\mathcal{I}_{m,C}\}$. Let us denote the $c$th RP as $\mathcal{I}_{m,c}=\{\mathcal{I}_{m,c}(0),\ldots,\mathcal{I}_{m,c}(K-1)\}$, where $\mathcal{I}_{m,c}(k)\in\mathbb{Z}_{+}^{G}$ for $k=0,\ldots,K-1$, and the corresponding activated RIS element index set is given by $\bar{\mathcal{I}}_{m,c}=\{\bar{\mathcal{I}}_{m,c}(0),\ldots,\bar{\mathcal{I}}_{m,c}(L_A-1)\}$, where $\mathcal{I}_{m,c}(l_A)\in\mathbb{Z}_{+}^{L}$ for $l_A=0,\ldots,L_A-1$. Moreover, the transmitted symbols of $U$ UE are generated based on a $Q$-ary normalized {quadrature amplitude modulation (QAM)/phase-shift keying (PSK)} constellation $\mathcal{B}=\{b_1,\ldots,b_Q\}$, yielding $L_2=U\log_2 Q$ bits. We consider a quasi-static block flat-fading channel model having a $\tau_c$ symbol-length coherence interval, while the length of the uplink channel estimation block is $\tau_p$ symbols. Consequently, the remaining $\tau_u=\tau_c-\tau_p$ symbols are utilized for uplink data transmission. It should be noted that the uplink SE at the UE side is only associated with the UE bit component $L_2$, yielding the sum SE as $\eta^{\text{SE}}=\sum_{u=1}^U\eta^{\text{SE}}_u=(\tau_u/\tau_c)\log_2 \left(1+\delta_u\right)$, where $\delta_u$ denotes uplink signal-to-interference-and-noise ratio (SINR) of UE $u$. On the other hand, the RIS bit component $L_1$ is taken into account when we calculate the amount of information contained at the AP side, as formulated in \eqref{eq3-2} and \eqref{eq3-3}. {Moreover, we emphasize that since the RIS RP is associated with the RIS-reflected channels, the bit sequence $L_1$ is only sent at the beginning of each coherence interval. This implies that the RP remains fixed during the coherence interval, and the RIS elements have to switch across the ON/OFF state, when a new coherence interval commences. Since the RIS controller has to detect the bit sequence $L_1$ to obtain the RIS RP, the hardware complexity of our proposed RPM-RIS-assisted CF-mMIMO is higher than that of the conventional schemes.}\footnote{{For the typical maximum likelihood detector, the complexity is on the order of $\mathcal{O}(2^{L_1})$. There are also other low-complexity detectors that can be considered, such as MMSE and approximate message passing (AMP) detectors \cite{9507331,10183832}.}}
\subsection{Channel Model}
The $l_A$th phase shift of the $m$-th RIS can be formulated as $\theta_{l_A,m}\in[-\pi,\pi],\forall m,l_A$. Therefore, the $m$-th RIS phase shift matrix can be expressed as $\pmb{\Phi}_m=\diag\{e^{i\theta_{1,m}},\ldots,e^{i\theta_{L_A,m}}\}\in\mathbb{C}^{L_A\times L_A}$. The channel matrix between the AP $m$ and the $m$-th RIS can be expressed as ${\pmb{G}}_m\in\mathbb{C}^{J\times L}$, while $\pmb{z}_{mu}\in\mathbb{C}^{L}$ denotes the channel vector between the $m$-th RIS and the UE $u$. We consider an urban environment having a multi-scatterer distribution, such that there is no direct path in the UE-AP link channel. Consequently, the channel between the AP $m$ and the UE $u$ is realistically modeled via the Rayleigh fading distribution, which is given as $\pmb{f}_{mu}\sim\mathcal{CN}(\pmb{0},\pmb{R}_{mu}), \forall m,u$, where $\pmb{R}_{mu}\in\mathbb{C}^{J\times J}$ denotes the corresponding spatial correlation matrix, and $\beta_{mu}=\tr(\pmb{R}_{mu})/J$ is the large-scale fading coefficient. Moreover, we consider the scenario that the UE-RIS and RIS-AP channels are modeled using the Rician distribution with high line-of-sight (LoS) probability. Explicitly, the UE-RIS and RIS-AP channels can be respectively formulated as
\begin{align}\label{eq: channel_reflect}
	&\pmb{z}_{mu}=\sqrt{\frac{\xi_{mu}}{\iota_{mu}+1}}\left(\sqrt{\iota_{mu}}\bar{\pmb{z}}_{mu}+\tilde{\pmb{z}}_{mu}\right),\nonumber\\
	 &\pmb{G}_m=\sqrt{\frac{\alpha_{m}}{\kappa_{m}+1}}\left(\sqrt{\kappa_{m}}\bar{\pmb{G}}_m+\tilde{\pmb{G}}_m\right),
\end{align}
where $\xi_{mu}$ and $\alpha_{m}$ are the path-loss coefficients, while $\iota_{mu}$ and $\kappa_m$ denote the Rician coefficients, respectively. Moreover, $\bar{\pmb{z}}_{mu}\in\mathbb{C}^{L_A}$ and $\bar{\pmb{G}}_m\in\mathbb{C}^{J\times L_A}$ represent the deterministic LoS components, while the non-LoS (NLoS) components can be formulated as $\tilde{\pmb{z}}_{mu}\sim\mathcal{CN}(\pmb{0},\tilde{\pmb{R}}_{mu})$ and $\text{vec}(\tilde{\pmb{G}}_m)\sim\mathcal{CN}(\pmb{0},\tilde{\pmb{R}}_{m})$. Moreover, $\tilde{\pmb{R}}_{mu}\in\mathbb{C}^{L_A\times L_A}$ and $\tilde{\pmb{R}}_{m}\in\mathbb{C}^{JL_A\times JL_A}$ denote the covariance and full correlation matrices, which can be respectively formulated as $\tilde{\pmb{R}}_{mu}=\pmb{R}_{\text{RIS},m}=d_H d_V\pmb{R}_m$ and $\tilde{\pmb{R}}_{m}=(\pmb{R}^T_{\text{AP},m}\otimes\pmb{R}_{\text{RIS},m})/JL_A$ \cite{10167480}, where $\pmb{R}_{\text{AP},m}$ and $\pmb{R}_{\text{RIS},m}$ denote the correlation matrices at the AP side \cite{hoydis2013massive} and the RIS side, respectively, while $d_H$ and $d_V$ denote the horizontal width and the vertical height of each RIS element. {Moreover, the RIS spatial correlation matrix $\pmb{R}_m\in\mathbb{C}^{L_A\times L_A}$ can be obtained based on \cite{bjornson2020rayleigh}.} For the LoS components, we consider that the AP and RIS have a $(J\times 1)$-uniform linear array (ULA) and $(\sqrt{L}\times\sqrt{L})$-uniform squared planar array (USPA), respectively. Therefore, the LoS components can be expressed as $\bar{\pmb{z}}_{mu}=\pmb{a}_{L_A}(\varphi_{mu}^a,\varphi_{mu}^e)$ and $\bar{\pmb{G}}_m=\pmb{a}_J(\upsilon_m^a,\upsilon_m^e)\pmb{a}_{L_A}(\bar{\varphi}_{m}^a,\bar{\varphi}_{m}^e)^H$ \cite{9973349}, where $\varphi_{mu}^a$ and $\varphi_{mu}^e$ are the azimuth and elevation angles of arrival (AoA) of the signal from the UE $u$ to RIS $m$. Furthermore, $\upsilon_m^a$ and $\upsilon_m^e$ denote the azimuth and elevation angles of departure (AoD) from the $m$-th RIS to the $m$-th AP, while $\bar{\varphi}_{m}^a$ as well as $\bar{\varphi}_{m}^e$ represent the azimuth and elevation AoA at the $m$-th AP, respectively. Moreover, {the elements of the steering vectors $\pmb{a}_{L_A}(\chi^a,\chi^e)\in\mathbb{C}^{L_A}$ and $\pmb{a}_J(\chi^a,\chi^e)\in\mathbb{C}^{J}$ can be obtained based on \cite{10326460}.} Given the activated RIS block index subset $\mathcal{I}_c$ and the $m$-th phase shift matrix, the RIS-assisted \emph{aggregated channel} spanning from the $u$-th UE to AP $m$ can be formulated as
\begin{align}\label{eq: channel}
	\pmb{h}_{mu}&=\pmb{f}_{mu}+{\pmb{G}}_m\pmb{\Phi}_m\pmb{z}_{mu}\triangleq \bar{\pmb{h}}_{mu}+\tilde{\pmb{h}}_{mu},
\end{align}
where we have $b_{mu}\triangleq {\frac{\alpha_m\xi_{mu}}{(\kappa_m+1)(\iota_{mu}+1)}}$, the LoS component is given by $\bar{\pmb{h}}_{mu}=\sqrt{b_{mu}\kappa_m\iota_{mu}}\bar{\pmb{G}}_m\pmb{\Phi}_m\bar{\pmb{z}}_{mu}$, the NLoS components are formulated as $\tilde{\pmb{h}}_{mu}^{1}=\sqrt{b_{mu}\kappa_m}\bar{\pmb{G}}_m\pmb{\Phi}_m\tilde{\pmb{z}}_{mu}$, $\tilde{\pmb{h}}_{mu}^{2}=\sqrt{b_{mu}\iota_m}\tilde{\pmb{G}}_m\pmb{\Phi}_m\bar{\pmb{z}}_{mu}$ and $\tilde{\pmb{h}}_{mu}^{3}=\sqrt{b_{mu}}\tilde{\pmb{G}}_m\pmb{\Phi}_m\tilde{\pmb{z}}_{mu}$, while $\bar{\pmb{h}}_{mu}$ and $\tilde{\pmb{h}}_{mu}\triangleq\sum_{c=1}^{3}\tilde{\pmb{h}}_{mu}^{c}+\pmb{f}_{mu}$ represent the LoS and NLoS components of the aggregated channel, respectively. We assume that the LoS component $\bar{\pmb{h}}_{mu}$ is known at all the APs \cite{ozdogan2019performance}. Moreover, we have the following statistical result of the unknown NLoS component $\tilde{\pmb{h}}_{mu}$.

\emph{Proposition 1: The covariance matrix of the NLoS component $\tilde{\pmb{h}}_{mu}$ of \eqref{eq: channel} can be formulated as}
\begin{align}\label{eq: NLoS covariance}
	\pmb{R}_{mu}^h\triangleq&\mathbb{E}\left\{\tilde{\pmb{h}}_{mu}\tilde{\pmb{h}}_{mu}^H\right\}=b_{mu}\kappa_m\bar{\pmb{G}}_m\pmb{\Phi}_m\tilde{\pmb{R}}_{mu}\pmb{\Phi}_m^H\bar{\pmb{G}}_m^H\nonumber\\
	&+\pmb{R}_{mu}+\pmb{\Pi}_{mu}+\pmb{\Xi}_{mu},
\end{align}
where $\pmb{\Pi}_{mu}\triangleq\mathbb{E}\left\{\tilde{\pmb{R}}_{mu,2}^h\right\}$, $\pmb{\Xi}_{mu}\triangleq\mathbb{E}\left\{\tilde{\pmb{R}}_{mu,3}^h\right\}$, while $\tilde{\pmb{R}}_{mu,2}^h$ and $\tilde{\pmb{R}}_{mu,3}^h$ are the correlation matrices of $\tilde{\pmb{h}}_{mu}^{2}$ and $\tilde{\pmb{h}}_{mu}^{3}$, respectively.

\emph{Proof:} The proof is offered in Appendix \ref{appendix2}. \hfill $\blacksquare$
\subsection{Uplink Channel Estimation}
By utilizing $\tau_p$ pilot symbols $\{\pmb{\phi}_{1},\ldots,\pmb{\phi}_{\tau_p}\}$ with $||\pmb{\phi}_t||^2=\tau_p$ transmitted by all UEs where $t\in\{1,\ldots,\tau_p\}$, we carry out the channel estimations independently, where we have $\tau_p<U$ and assume that $\tau_p$ pilot symbols are shared by $U$ UEs. Let $\mathcal{P}_u\subset\{1,\ldots,U\}$ represent the UE index set that harnesses the same pilot symbol as the UE $u$, including the $u$-th UE itself. More specifically, the pilot symbol allocated to UE $u$ is given by $\pmb{\phi}_{t_u}\in\mathbb{C}^{\tau_p}$ with $t_u\in\{1,\ldots,\tau_p\}$. Given UE $u$ and $\mathcal{P}_u$, the pilot reuse pattern can be formulated as $\pmb{\phi}_{t_{k}}^T\pmb{\phi}_{t_u}^*=\tau_p,  k\in\mathcal{P}_u$ and $\pmb{\phi}_{t_{k}}^T\pmb{\phi}_{t_u}^*=0,  k\notin\mathcal{P}_u$. The symbol received at AP $m$ can be formulated as $\pmb{Y}_m^p=\sum_{k=1}^{U}\sqrt{p_k}\pmb{h}_{mk}\pmb{\phi}_{t_k}^T+\pmb{N}_m$, where $p_k\geq 0$ represents the normalized SNR per pilot symbol of the $k$th UE during channel estimation, while the elements of the noise term $\pmb{N}_m\in\mathbb{C}^{J\times \tau_p}$ are independent and identically distributed (i.i.d.) Gaussian random variables (RVs) with zero mean and unit variance. Consequently, the pilot-projected-version of the received training signal $\pmb{y}_{mu}^p=\pmb{Y}_m^p\pmb{\phi}_{t_u}^*\in\mathbb{C}^J$ can be obtained as
\begin{align}\label{eq: pilot project}
	\pmb{y}_{mu}^p=\sqrt{p_u}\tau_p\pmb{h}_{mu}+\sum_{k\in\mathcal{P}_u\setminus\{u\}}\sqrt{p_k}\tau_p\pmb{h}_{mk}+\pmb{n}_{mu},
\end{align}
where $\pmb{n}_{mu}=\pmb{N}_m\pmb{\phi}_{t_u}^*/\sqrt{\tau_p}\sim\mathcal{CN}(\pmb{0},\tau_p\pmb{I}_J)$. Based on \eqref{eq: pilot project}, and upon invoking the linear MMSE channel estimation technique \cite{bjornson2019making}, the estimated aggregated channel can be formulated as
\begin{align}\label{eq: channel estimate}
	\hat{\pmb{h}}_{mu}=\bar{\pmb{h}}_{mu}+\sqrt{p_u}\pmb{R}^h_{mu}\pmb{\Psi}_{mu}^{-1}(\pmb{y}^p_{mu}-\bar{\pmb{y}}^p_{mu}),
\end{align}
where $\pmb{\Psi}_{mu}=\sum_{k\in\mathcal{P}_u}{p_k}\tau_p\pmb{R}^h_{mk}+\pmb{I}_J$ and $\bar{\pmb{y}}^p_{mu}=\sum_{k\in\mathcal{P}_u}\sqrt{p_k}\tau_p\bar{\pmb{h}}_{mk}$. The expectation and covariance matrices of the channel estimate $\hat{\pmb{h}}_{mu}$ and the estimation error $\tilde{\pmb{h}}_{mu}={\pmb{h}}_{mu}-\hat{\pmb{h}}_{mu}$ can be respectively derived as $\mathbb{E}\{\hat{\pmb{h}}_{mu}\}=\bar{\pmb{h}}_{mu}$, $\text{Cov}\{\hat{\pmb{h}}_{mu}\}=\bar{\pmb{C}}_{mu}=p_u\tau_p\pmb{\Gamma}_{mu}$ and $\mathbb{E}\{\tilde{\pmb{h}}_{mu}\}=\pmb{0}$, $\text{Cov}\{\tilde{\pmb{h}}_{mu}\}=\pmb{\Lambda}_{mu}$, where we have
\begin{align}\label{eq: channel estimate matrix}
\pmb{\Gamma}_{mu}\triangleq\pmb{R}_{mu}^h\pmb{\Psi}_{mu}^{-1}\pmb{R}_{mu}^h,\quad \pmb{\Lambda}_{mu}\triangleq\pmb{R}_{mu}^h-\bar{\pmb{C}}_{mu}.
\end{align}
It should be noted that $\hat{\pmb{h}}_{mu}$ and $\tilde{\pmb{h}}_{mu}={\pmb{h}}_{mu}-\hat{\pmb{h}}_{mu}$ are uncorrelated.
\section{Uplink Data Transmission and Spectral Efficiency}\label{Section 3}
\subsection{Uplink Data Transmission Phase}
Upon denoting the symbol transmitted by the $u$-th user as $s_u=\sqrt{\bar{p}_u}\bar{s}_u$, where $\bar{p}_u$ represents the uplink power control coefficient, and $\bar{s}_u$ is generated based on a $Q$-ary normalized QAM/PSK constellation $\mathcal{B}=\{b_1,\ldots,b_Q\}$ with $\mathbb{E}\{|\bar{s}_u|^2\}=1$. The $J$-dimensional signal received at the $m$-th AP is given by
\begin{align}\label{eq: UL input-ouput}
\pmb{y}_m&=\sum_{u=1}^{U}\pmb{h}_{mu}s_u+\pmb{n}_m=\pmb{H}_m\pmb{P}^{1/2}\bar{\pmb{s}}+\pmb{n}_m,
\end{align}
where $\pmb{P}^{1/2}=\diag\{\bar{\pmb{p}}\}$ with $\bar{\pmb{p}}=[\sqrt{\bar{p}_1},\ldots,\sqrt{\bar{p}_U}]^T$ denoting the uplink power control matrix, $\bar{\pmb{s}}=[\bar{s}_1,\ldots,\bar{s}_U]^T$ is the symbol vector, $\pmb{H}_m=[\pmb{h}_{m1},\ldots,\pmb{h}_{mU}]\in\mathbb{C}^{J\times U}$ while ${{\pmb{n}}_m} \sim \mathcal{C}\mathcal{N}\left( {\pmb{0},{\sigma ^2}{{\pmb{I}}_J}} \right)$ represents the noise term. We assume that each AP is capable of processing the uplink received signal locally by applying a specific combining vector $\pmb{v}_{mu}\in\mathbb{C}^J$ to $\pmb{y}_m$ \cite{bjornson2019making,10167480}. Hence, the local symbol estimates $\tilde{s}_{mu}=\pmb{v}^H_{mu}\pmb{y}_m$ can be formulated as
\begin{align}\label{eq: local estimate}	\tilde{s}_{mu}=\pmb{v}^H_{mu}\pmb{h}_{mu}s_u+\sum_{k\neq u}^U\pmb{v}^H_{mu}\pmb{h}_{mk}s_k+\pmb{v}^H_{mu}\pmb{n}_m,
\end{align}
where any combiner can be exploited in \eqref{eq: local estimate}. In this paper, we first consider the simple MR combining scheme associated with $\pmb{v}_{mu}=\hat{\pmb{h}}_{mu}$. Moreover, we invoke the L-MMSE combiner $\pmb{v}_{mu}=\bar{p}_u\left[\sum_{k=1}^{U}\bar{p}_k\left(\hat{\pmb{h}}_{mk}\hat{\pmb{h}}_{mk}^H+\pmb{\Lambda}_{mk}\right)+\sigma^2\pmb{I}_M\right]^{-1}\hat{\pmb{h}}_{mu}$ to attain the minimum $\text{MSE}=\mathbb{E}\left\{\lVert{s}_u-\pmb{v}^H_{mu}\pmb{y}_m\rVert^2|\hat{\pmb{h}}_{mk}\right\}$. Then, we exploit the LSFD scheme to mitigate the interference \cite{bjornson2019making}. All the local symbol estimates $\{\tilde{s}_{mu}, \forall m\}$ are sent to the CPU and are linearly combined by applying the weighting factors $c_{mu}$ for $m=1,\ldots,M$, yielding the final estimate $\hat{s}_u=\sum_{m=1}^M c_{mu}^*\tilde{s}_{mu}$ as
\begin{align}\label{eq: final estimate}
{{\hat s}_u}= {\pmb{c}}_u^H{{\pmb{g}}_{uu}}{s_u} + \sum\limits_{k\neq u}^U {{\pmb{c}}_u^H{{\pmb{g}}_{uk}}{s_k}}+{{{n}}_u},
\end{align}
where we have $\pmb{g}_{uk}=[\pmb{v}^H_{1u}\pmb{h}_{1k},\ldots,\pmb{v}^H_{Mu}\pmb{h}_{Mk}]^T\in\mathbb{C}^{M}$ and $\pmb{c}_u=[c_{1u},\ldots,c_{Mu}]^T\in\mathbb{C}^{M}$ denotes the weighting factor vector, ${{{n}}_u}=\sum\limits_{m = 1}^M {c_{mu}^ * } {\pmb{v}}_{mu}^H{{\pmb{n}}_m}$, while the equivalent channel gain can be expressed as $\{\pmb{c}^H_{u}\pmb{g}_{uk},k=1,\ldots,U\}$.

\subsection{Uplink Spectral Efficiency}
We assume that each UE is capable of obtaining only the channel statistics but not the specific CSI. Upon exploiting the use-and-then-forget bounding method \cite{7827017}, the uplink SE associated with UE $u$ (in bit/s/Hz) can be expressed based on \eqref{eq: final estimate} as
\begin{align}\label{eq: SEu}
	\eta^{\text{SE}}_u=(\tau_u/\tau_c)\log_2 \left(1+\delta_u\right),
\end{align}
where $\delta_u$ is the effective uplink SINR of UE $u$, defined as follows
{\begin{align}\label{eq: SINR monte carlo}
	\delta_u&=\frac{\bar{p}_u|\pmb{c}_u^H\mathbb{E}\{\pmb{g}_{uu}\}|^2}{\sum\limits_{k=1}^U\bar{p}_k\mathbb{E}\{|\pmb{c}_u^H\pmb{g}_{uk}|^2\}-\bar{p}_u|\pmb{c}_u^H\mathbb{E}\{\pmb{g}_{uu}\}|^2+\sigma^2\pmb{c}_u^H\pmb{V}_u\pmb{c}_u}\nonumber\\
	&=\frac{\bar{p}_u|\pmb{c}_u^H\mathbb{E}\{\pmb{g}_{uu}\}|^2}{\pmb{c}^H_u\left(\sum\limits_{k=1}^U \bar{p}_k\pmb{\Omega}_{uk}-\bar{p}_u\pmb{G}_u+\sigma^2\pmb{V}_u\right)\pmb{c}_u},
\end{align}
where $\pmb{G}_u=\mathbb{E}\{\pmb{g}_{uu}\}\mathbb{E}\{\pmb{g}_{uu}^H\}$} and $\pmb{\Omega}_{uk}\in\mathbb{C}^{M\times M}$ represents the inter-user interference introduced by pilot contamination and imperfect channel estimation, whose elements can be expressed as $\mathbb{E}\{\pmb{v}_{mu}^H\pmb{h}_{mk}\pmb{h}_{m'k}^H\pmb{v}_{m'u}\},\forall m,m'$, and we have $\pmb{V}_u={\diag}\left\{\mathbb{E}\left[||\pmb{v}_{1u}||^2\right],\ldots,\mathbb{E}\left[||\pmb{v}_{Mu}||^2\right]\right\}$.

\emph{Proposition 2: Upon exploiting the MR combining scheme associated with $\pmb{v}_{mu}=\hat{\pmb{h}}_{mu}$ and the LSFD technique, the uplink SINR of UE $u$ can be expressed in closed-form as
{\begin{align}\label{eq: SINR closed form}
	\delta_u=\frac{\bar{p}_u|{\tr}(\pmb{C}_u^H\pmb{\Upsilon}_{u})|^2}{\sum\limits_{k=1}^U\bar{p}_k{\tr}(\bar{\pmb{C}}_{uk})+\sum\limits_{k\in\mathcal{P}_u\setminus\{u\}}\bar{p}_k{L}_{uk}+{\tr}(\pmb{D}_u)},
\end{align}
where we have $\bar{\pmb{C}}_{uk}=\pmb{C}_u^H\pmb{T}_{uk}\pmb{C}_u$} and $\pmb{D}_u=\pmb{C}_u^H(\sigma^2\pmb{\Upsilon}_{u}-\bar{p}_u\pmb{\Delta}_u^2)\pmb{C}_u^H$. In \eqref{eq: SINR closed form}, we specifically denote $\pmb{C}_u={\diag}\{\pmb{c}_u\}\in\mathbb{C}^{M\times M}$ and $\pmb{\Upsilon}_u={\diag}\{\bar{\pmb{\xi}}_u\}\in\mathbb{C}^{M\times M}$, where $\bar{\pmb{\xi}}_u=[\bar{\xi}_{1u},\ldots,\bar{\xi}_{Mu}]^T$ with $\bar{\xi}_{mu}=\bar{\pmb{h}}_{mu}^H\bar{\pmb{h}}_{mu}+p_u\tau_p{\tr}(\pmb{\Gamma}_{mu})$ associated with the desired signal. Moreover, the non-coherent interference can be formulated as $\pmb{T}_{uk}={\diag}\{\pmb{\mu}_{uk}\}\in\mathbb{C}^{M\times M}$ and $\pmb{\mu}_{uk}=[\mu_{uk,1},\ldots,\mu_{uk,M}]^T$, where
\begin{align}
	\mu_{uk,m}=&p_k\tau_p{\tr}(\pmb{\Gamma}_{mu}\pmb{R}^h_{mk})+\bar{\pmb{h}}^H_{mu}\pmb{R}^h_{mk}\bar{\pmb{h}}_{mu}\nonumber\\
	&+p_k\tau_p\bar{\pmb{h}}^H_{mk}\pmb{\Gamma}_{mu}\bar{\pmb{h}}_{mk}+|\bar{\pmb{h}}^H_{mu}\bar{\pmb{h}}_{mk}|^2.
\end{align}
Furthermore, the coherent interference ${L}_{uk}$ can be expressed by ${L}_{uk}=p_u p_k\tau_p^2|{\tr}(\pmb{C}_u\pmb{W}_{uk})|^2$ and $\pmb{W}_{uk}={\diag}\{\pmb{w}_{uk}\}\in\mathbb{C}^{M\times M}$, where $\pmb{w}_{uk}=[w_{uk,1},\ldots,w_{uk,M}]^T$ with $w_{uk,m}={\tr}(\pmb{R}_{mk}^h\pmb{\Psi}^{-1}_{mu}\pmb{R}_{mu}^h)$, and $\pmb{\Delta}_u={\diag}\left\{||\bar{\pmb{h}}_{1u}||^2,\ldots,||\bar{\pmb{h}}_{Mu}||^2\right\}$.}

\emph{Proof: }The proof is provided in Appendix \ref{appendix3}. \hfill $\blacksquare$

It should be noted that \eqref{eq: SINR monte carlo} is in a generalized Rayleigh quotient form. Therefore, based on the technique in \cite{bjornson2019making}, the optimal LSFD weighting vector $\pmb{c}_u^\text{opt}$ optimized for maximizing the SINR given in \eqref{eq: SINR monte carlo} is
\begin{align}\label{eq:optweighting}
	\pmb{c}_u^\text{opt}=\left(\sum_{k=1}^U \bar{p}_k\pmb{\Omega}_{uk}-\bar{p}_u\pmb{G}_u+\sigma^2\pmb{V}_u\right)^{-1}\mathbb{E}\{\pmb{g}_{uu}\},
\end{align}
yielding the corresponding maximum SINR as $\delta_u^\text{opt}=\bar{p}_u\mathbb{E}\{\pmb{g}^H_{uu}\}\pmb{c}_u^\text{opt}$. Consequently, the sum SE can be expressed as $\eta^{\text{SE}}=\sum_{u=1}^U\eta^{\text{SE}}_u$.
\section{Power Consumption and Energy Efficiency}\label{Section 4}
Based on the standard random coding theory of \cite{cover1999elements}, the transmitted symbol vector $\pmb{\bar{s}}$ can be modeled as $\bar{\pmb{s}}\sim\mathcal{CN}(\pmb{0},\pmb{I}_U)$. Therefore, based on \eqref{eq: UL input-ouput}, given the effective channel matrix $\pmb{H}_m$ and the uplink power control matrix ${\pmb{P}}$, it can be observed that the received signal follows the distribution of $\pmb{y}_m\sim\mathcal{CN}(\pmb{0},\pmb{H}_m\pmb{P}\pmb{H}_m^H+\sigma^2\pmb{I}_J)$. Therefore, the conditional entropy of $\pmb{y}_m$ and the differential entropy $\pmb{n}_m$ can be respectively formulated as \cite{cover1999elements}
\begin{align}\label{eq3-2}
	\mathcal{H}(\pmb{y}_m|\pmb{H}_m)=&\log_2 \{\det[\pi e(\pmb{H}_m\pmb{P}\pmb{H}_m^H+\sigma^2\pmb{I}_J)]\},\nonumber\\
	 \mathcal{H}(\pmb{n}_m)=&\log_2 \{\det(\pi e\sigma^2\pmb{I}_J)\}.
\end{align}
Given the RP and AP index $m$, the channel matrix $\pmb{H}_m$ is seen to be a function of the RIS phase shift matrix $\pmb{\Phi}_m$. Therefore, based on the mutual information between $\pmb{H}_m\pmb{P}^{1/2}\bar{\pmb{s}}$ and $\pmb{y}_m$, the amount of information contained in $\pmb{y}_m$ measured in (bit/s/Hz) can be formulated as
\begin{align}\label{eq3-3}
	\bar{C}_m&=\mathbb{E}_{\pmb{H}_m}\{I(\pmb{y}_m;\pmb{H}_m\pmb{P}^{1/2}\bar{\pmb{s}}|\pmb{H}_m)\}\nonumber\\
	&=\mathbb{E}_{\pmb{H}_m}\{\mathcal{H}(\pmb{y}_m|\pmb{H}_m)\}-\mathcal{H}(\pmb{n}_m)\nonumber\\
	&=\mathbb{E}_{\pmb{H}_m}\left\{\log_2\det \left(\frac{\pmb{H}_m\pmb{P}\pmb{H}_m^H}{\sigma^2}+\pmb{I}_J\right)\right\}.
\end{align}

{Consequently, the total uplink power consumption can be formulated as $P^{\text{tot}}=\sum_{m=1}^{M}P_m^{\text{AP}}+\sum_{m=1}^{M}P^{\text{BH}}_m+\sum_{m=1}^{M}P^{\text{RIS}}_m+\sum_{u=1}^U P_u^{\text{UE}}$\cite{8097026,huang2019reconfigurable,femenias2021swipt}, where $P_m^{\text{AP}}$ and $P^{\text{BH}}_m$ represent the $m$-th AP's power consumption and the corresponding backhaul link, respectively. Furthermore, $P_m^{\text{RIS}}$ and $P_u^{\text{UE}}$ represent the power consumption of the $m$-th RIS and $u$-th UE, respectively. Also, $P_m^{\text{AP}}$ can be expressed as $P_m^{\text{AP}}=B\frac{\tau_u}{\tau_c}\bar{C}_m\varrho^{\text{AP}}_m+P_m^{\text{AP,fix}}+JP_{m}^{\text{AP,a}}$, where $\varrho^{\text{AP}}_m$ denotes the traffic-dependent power coefficient measured in (Watt/bit/s), $P_m^{\text{AP,fix}}$ is the fixed AP power consumption related to the traffic load, while $P_{m}^{\text{AP,a}}$ is the traffic-agnostic uplink power consumed by each antenna of the $m$-th AP. Moreover, we have $P^{\text{BH}}_m=B\frac{\tau_u}{\tau_c}\bar{C}_m\varrho^{\text{BH}}_m+P^{\text{BH,fix}}_m$, with bandwidth $B$, while $\varrho^{\text{BH}}_m$ denotes the traffic-dependent power coefficient measured in (Watt/bit/s), and $P^{\text{BH,fix}}_m$ represents the fixed power dissipated by the $m$-th backhaul link. Then, the RIS power consumption is given by $P^{\text{RIS}}_m=L_A P(b),\forall m$, where $P(b)$ is the power consumed by each RIS element having $b$-bit resolution \cite{huang2019reconfigurable}. Finally, we have $P_u^\text{UE}=\bar{p}_u/\alpha^{\text{UE}}+P_u^\text{UE,fix}$ with the power amplifier efficiency $\alpha^{\text{UE}}$ and UE-side fixed power consumption $P_u^\text{UE,fix}$. The total uplink power consumption can be expressed as
\begin{align}\label{eq: Ptot}
	P^{\text{tot}}=P^{\text{fix}}+\gamma\sum_{u=1}^U\bar{p}_u+P^{\text{RIS}}_m+ B\frac{\tau_u}{\tau_c}\sum_{m=1}^M\bar{C}_m\tilde{\varrho}_m,
\end{align}
where $P^\text{fix}$ denotes the fixed power consumption, $\tilde{\varrho}_m=\varrho^{\text{AP}}_m+\varrho^{\text{BH}}_m$,} and the total EE measured in (bit/Joule) can be formulated as $\eta^{\text{EE}}={B\eta^{\text{SE}}}/{P^\text{total}}$ \cite{8097026}.

{The static power consumption of RIS includes the adjustable electronic circuit and the number of control-signal-related components \cite{10480438}. For the RPM-RIS-assisted CF-mMIMO systems, the power dissipated by the electronic circuit is higher than that of the conventional RIS-assisted CF-mMIMO, since the ON/OFF states of RIS elements are switched at the beginning of each coherence interval. However, only a fraction of the RIS elements are activated in the proposed RPM-RIS-assisted CF-mMIMO systems, yielding lower power consumption than the RIS-assisted CF-mMIMO. Therefore, we still exploit the conventional power consumption model of \eqref{eq: Ptot} in this paper. A detailed power consumption analysis of our RPM-RIS-assisted CF-mMIMO will be carried out in our future work.}
\section{RIS Phase Shift Design for Uplink Energy Efficiency Maximization}\label{Section 5}
\subsection{Problem Statement}
Upon denoting the collected set of RIS phase shift matrices as $\pmb{\Phi}=\{\pmb{\Phi}_1,\ldots,\pmb{\Phi}_M\}$, hence the optimization problem of maximizing the total EE can be formulated as
\begin{subequations}\label{eq: opt1}
	\begin{empheq}[left={({P_0}) :}\empheqlbrace]{align}
 \max_{\pmb{\Phi}}\ &\eta^{\text{EE}}\label{eq: obj}\\
	\text{s.t.}\ & \eta^{\text{SE}}_u\geq \eta_{u,\text{min}}^\text{EE},\forall u,\label{eq: QoS1}\\
	& -\pi\leq\theta_{l_A,m}\leq \pi,\forall l_A,m.\label{eq: ps1}
\end{empheq}
\end{subequations}
It can be readily shown that the optimization problem $(P_0)$ that the objective function \eqref{eq: obj} and the constraint of \eqref{eq: QoS1} are non-convex. {The EE optimization problem of existing works is formulated based on the {instantaneous} SINR \cite{10290997,10175060,9363171,10035459}. By contrast, our corresponding optimization problem $(P_0)$ is conceived based on our new closed-form expression of the ergodic SINR. Consequently, it can be observed from \eqref{eq: SINR closed form} and \eqref{eq: Ptot} that the expression of $\eta^\text{EE}$ is complex and the phase shift matrices $\pmb{\Phi}_m$ for $m=1,\ldots,M$ are highly coupled. Hence, it is challenging to derive the globally optimal solution of \eqref{eq: opt1} upon utilizing conventional optimization algorithms, such as majorization-minimization (MM), gradient ascent-based algorithms and successive convex approximation (SCA) algorithms. Hence we opted for the heuristic PSO technique.} Therefore, we propose a CSA-PSO scheme for solving the phase shift design optimization problem shown in \eqref{eq: opt1}.
\subsection{Phase Shift Optimization Using CSA-PSO}
We assume that the CSA-PSO population $\mathcal{Q}$ includes $I$ individuals.
Based on the system model shown in Section \ref{Section 2}, there are a total of $ML_A$ phase shift elements. yielding $\pmb{\theta}=\{\pmb{\theta}_1,\ldots,\pmb{\theta}_M\}$ associated with $\pmb{\theta}_m\in\mathbb{C}^{L_A},\forall m$. Hence, the $i$th initial individual can be expressed as $\pmb{\theta}_i^{0}=\{\theta_{i}^0(1),\ldots,\theta_{i}^0(ML_A)\}$, where each element is in the range of $[-\pi,\pi]$, and can be formulated as $\theta^0_{i}(d)=-\pi+2\pi\kappa_{i}(d),\forall i,d,
$ for $d=1,\ldots,ML_A$ and $i=1,\ldots,I$. Moreover, $\kappa_{i}(d)\in[0,1]$ is the chaotic sequence that can be obtained via the Logistic function \cite{caponetto2003chaotic}, which can be generated as $\kappa_{i}(d+1)=\tilde{\mu}\kappa_{i}(d)[1-\kappa_{i}(d)]$, where $\tilde{\mu}=4$ and $\kappa_{i}(1)\notin\{0,0.25,0.5,0.75,1\}$. During the $t$th iteration, all the $I$ populations can be updated in parallel. According to the fitness value that each particle has achieved, the CSA-PSO algorithm registers both the local and global optimal particles, which are denoted as $\pmb{\theta}_{i,\text{pbest}}^{t}$ and $\pmb{\theta}_{\text{gbest}}^{t}$, respectively. We propose an update scheme of the velocities of each particle, which can be formulated as $\pmb{v}^{t+1}_{i}=\omega^{t}\pmb{v}^{t}_{i}+c_{1} r_1\left(\pmb{\theta}_{i,\text{pbest}}^{t}-\pmb{\theta}_{i}^t\right)+c_{2} r_2\left(\pmb{\theta}_{\text{gbest}}^{t}-\pmb{\theta}_{i}^t\right)$, where $\omega^{t}$ denotes the inertial weight factor, $c_{1}$ and $c_{2}$ are the acceleration factors, while we have $r_k\sim\mathcal{U}(0,1)$ for $k=1,2$. Moreover, we have $c_1=c_2=1.496$ \cite{985692}. It should be noted that the velocity is limited by $[-v_{\text{max}},v_{\text{max}}]$. We propose adaptive strategies to adjust the inertial weight factor, yielding
\begin{align}\label{eq: weightf}
	\omega^{t}=\omega_{\text{min}}+(\omega_{\text{max}}-\omega_{\text{min}})\left(\frac{2}{1+e^{-5\zeta^t}}-1\right),
\end{align}
where $\omega_{\text{min}}$ and $\omega_{\text{max}}$ are the preset minimum and maximum values of the inertial weight factor, $\zeta^t=(T_\text{max}-t)/T_\text{max}$, where $T_{\text{max}}$ denotes the maximum number of iterations. {Consequently, the initial weight factor can be obtained when $t=0$, i.e., we have $\omega^0=\omega_{\text{min}}$.} The particles are updated as $\pmb{\theta}_{i}^{t+1}=\pmb{\theta}_{i}^t+\pmb{v}_{i}^{t+1}$. To further enhance the performance, we propose a mutation scheme, where the worst local particle $\pmb{\theta}^t_{\text{pworst}}$ is mutated as $\pmb{\theta}^t_{\text{pworst}}=\pmb{\theta}^t_{\text{gbest}}+\pmb{\theta}_\Delta^t$, where $\pmb{\theta}_\Delta^t\sim\mathcal{N}(\pmb{0},\sigma^2_t\pmb{I}_{ML_A})$ with $\sigma_t=2\pi(1-t/T_{\text{max}})$. Since the conventional PSO algorithm might get trapped in a local optimum, we check the fitness values of all the particles based on their history. If the fitness values remain unchanged for $T_\text{check}$ iterations, we reset the corresponding particles based on $\pmb{\theta}_{\text{gbest}}^{t}$ and $\pmb{\theta}_{i,\text{pbest}}^{t}$, yielding $\theta_{i}^{t+1}(d)\sim\mathcal{N}\left(\frac{\theta_{\text{gbest}}^t(d)-\theta_{i,\text{pbest}}^{t}(d)}{2},\left|\theta_{\text{gbest}}^t(d)-\theta_{i,\text{pbest}}^{t}(d)\right|\right)$ for $i\in\{1,\ldots,I\}$. After $T_\text{max}$ iterations, the best individual in a set of $\mathcal{Q}$ is selected as the optima. The proposed CSA-PSO algorithm is summarized in Algorithm \ref{alg1}.
\begin{algorithm}[htbp]
\footnotesize
\caption{CSA-PSO Algorithm}
\label{alg1}
\begin{algorithmic}[1]
    \Require
      $\kappa_i(d)$, $\tilde{\mu}$, $\omega^0$, $c_1$, $c_2$, $\omega_{\text{min}}$ and $\omega_{\text{max}}$.      
    \State \textbf{Preparation}: Set a maximum iterations number $T_{\text{max}}$, $t_{i,\text{check}}=0$, $n_i=0$ and $N_i=10$
    \State \textbf{Initialize} Compute $\pmb{\theta}_i^0$ using chaotic sequence.
    \While{$t=0<T_{\text{max}}{\land n_i<N_i}$}
    \State Calculate $(\eta_i^\text{EE})^t$.
    \State Obtain the local best particles $\pmb{\theta}_{i,\text{pbest}}^{t}$, global best particles $\pmb{\theta}_{\text{gbest}}^{t}$ and local worst particles $\pmb{\theta}_{\text{pworst}}^{t}$ based on $(\eta_i^\text{EE})^t$.
    \State Mutate the local worst particle as $\pmb{\theta}^t_{\text{pworst}}=\pmb{\theta}^t_{\text{gbest}}+\pmb{\theta}_\Delta^t$.
    \State $\omega^{t}=\omega_{\text{min}}+(\omega_{\text{max}}-\omega_{\text{min}})\left(\frac{2}{1+e^{-5\zeta^t}}-1\right)$.
    \State $\pmb{v}^{t+1}_{i}=\omega^{t}\pmb{v}^{t}_{i}+c_{1} r_1\left(\pmb{\theta}_{i,\text{pbest}}^{t}-\pmb{\theta}_{i}^t\right)+c_{2} r_2\left(\pmb{\theta}_{\text{gbest}}^{t}-\pmb{\theta}_{i}^t\right)$.
    \State Obtain the updated particles as $\pmb{\theta}_{i}^{t+1}=\pmb{\theta}_{i}^t+\pmb{v}_{i}^{t+1}$.
    \State \textbf{while} $t>0$
    \State \quad \quad\quad\textbf{if} $(\eta_i^\text{EE})^t=(\eta_i^\text{EE})^{t-1}$ \textbf{then}
    \State \quad\quad\quad\quad$t_{i,\text{check}}=t_{i,\text{check}}+1$.
    \State \quad\quad\quad\textbf{else} $t_{i,\text{check}}=0$.
    \State \quad \quad\quad\textbf{end if}
    \State \textbf{end while}
    \State \textbf{If} $t_{i,\text{check}}>T_{\text{check}}$
    \State\quad $\theta_{i}^{t+1}(d)\sim\mathcal{N}\left(\frac{\theta_{\text{gbest}}^t(d)-\theta_{i,\text{pbest}}^{t}(d)}{2},\left|\theta_{\text{gbest}}^t(d)-\theta_{i,\text{pbest}}^{t}(d)\right|\right)$.
    \State \textbf{end if}
    {\State \textbf{If} $|(\eta_i^\text{EE})^t-(\eta_i^\text{EE})^{t-1}|<\epsilon$}    
    {\State\quad $n_i=n_i+1$}
    {\State \textbf{end if}}
    \EndWhile
\State \textbf{return} $\pmb{\theta}_{\text{gbest}}^{T_{\text{max}}+1}$ and the corresponding EE value $\eta^{\text{EE}}$.
\end{algorithmic}
\end{algorithm}
\subsection{Complexity Analysis}\label{Section 3-3}
In the conventional PSO algorithm, the initial phase shifts $\pmb{\theta}_i^0$ are generated randomly. Moreover, the inertial factor is fixed \cite{985692}. Furthermore, the mutation and particle reset steps are saved. The complexity of phase shift initialization is on the order of $\mathcal{O}(ML_A I)$. During each iteration, obtaining the best local and global solutions by sorting the particles based on their EEs requires $\mathcal{O}(I\log I)$ operations. The complexity of calculating the velocities is given by $\mathcal{O}(ML_A I)$. Hence, we can readily show that the overall complexity of PSO is on the order of $\mathcal{O}[ML_A I+(I\log I+ML_A I)T_{\text{max}}]=\mathcal{O}[(I\log I+ML_A I)T_{\text{max}}]$.

Compared to the conventional PSO, our proposed CSA-PSO invokes chaotic sequences to generate the initial RIS phase shifts, while the complexity of each step is on the order of $\mathcal{O}(ML_AI)$. Additionally, the complexity of updating the inertial weight factor based on \eqref{eq: weightf} is given by $\mathcal{O}(T_\text{max})$. The complexity of the particle reset step is on the order of $\mathcal{O}(ML_AT_1)$, while the best case is associated with $t_{i,\text{check}}\leq T_{\text{check}}$ during the iterations, yielding $T_1=0$. The worst case occurs, when the particles are reset every $T_\text{check}$ times, and we have $T_1=\left\lfloor{T_\text{max}}/{T_\text{check}}\right\rfloor$. Finally, the complexity of the proposed CSA-PSO is given as $\mathcal{O}[ML_A I+ML_A I+(I\log I+ML_AI+1)T_{\text{max}}+ML_AT_1]=\mathcal{O}[(I\log I+ML_A I)T_{\text{max}}+ML_A(T_1+2I)]$. Since the complexity is dominated by $\mathcal{O}[(I\log I+ML_A I)T_{\text{max}}]$, it can be observed that the complexity of our CSA-PSO is only slightly higher compared to the PSO algorithm.
\vspace{-1em}
\section{Simulation Results}\label{Section 6}
In this section, we provide simulation results for characterizing the SE and EE of our proposed RPM-RIS-assisted CF-mMIMO system relying on our CSA-PSO phase shift optimization algorithm. For all simulations, we consider $M$ APs and $U$ UEs, which are independently and uniformly distributed inside a $1\times 1\ \text{km}^2$ area, where a wrap-around setup is emulated for avoiding the `desert island' edge effects. The heights of AP, RIS and UE are set as $12.5$ m, $30$ m and $1.5$ m, respectively. {Based on the micro-cellular COST 321 Walfish-Ikegami model, the path-loss (PL) coefficients of the NLoS component associated with the UE-AP, UE-RIS and RIS-AP links can be formulated as $\text{PL}_\text{NLoS}[\text{dB}]=-34.53-38\log_{10}\left(d/1 \text{m}\right)+F$ \cite{ozdogan2019performance,10167480,9737367},} where $d$ denotes the communication distance according to the scenario, where $F=\sqrt{\delta_f}a+\sqrt{1-\delta_f}b$ represents the shadow fading having the parameter $\delta_f$ \cite{7827017,ozdogan2019performance,9737367}. Moreover, $a\sim\mathcal{N}(0,\delta_\text{sf}^2)$ and $b\sim\mathcal{N}(0,\delta_\text{sf}^2)$ are independent random variables at the transmitter and receiver sides, respectively. Consider the UE-AP direct link as an example. The covariance functions can be formulated as $\mathbb{E}\{a_u a_{u'}\}=2^{-\frac{d_{uu'}}{d_\text{dc}}}$ and $\mathbb{E}\{b_m b_{m'}\}=2^{-\frac{d_{mm'}}{d_\text{dc}}}$, where $d_{uu'}$ and $d_{mm'}$ are the UE $u$-UE $u'$ and AP $m$-AP $m'$ distances, respectively, while $d_\text{dc}$ denotes the decorrelation distance \cite{7827017}. The Rician factors are formulated as $\iota_{mu}=10^{1.3-0.003d_{mu}}$ and $\kappa_m=10^{1.3-0.003d_{u}}$, where $d_{mu}$ and $d_{u}$ represent the distances of the UE $u$-RIS $m$ and RIS $m$-AP $m$ pairs, respectively. Here, we set $\delta_f=0.5$, $\delta_\text{sf}=8$ and $d_\text{dc}=100$ m \cite{ozdogan2019performance,9737367}, while we select $d_m=10$ m, bandwidth $B=20$ MHz and noise power $\sigma^2=-94$ dBm. The transmit power is $p_u=\bar{p}_u=200$ mW, $\forall u$, the length of each coherence block is $\tau_c=200$ where $\tau_p=2$ pilots are used in each coherence block, {which corresponds to a coherence bandwidth of $B_c=200$ KHz and a coherence time of $T_c=1$ ms.} Each RIS contains $L=64$ elements and it is uniformly divided into $G=4$ blocks with $d_H=d_V=\lambda/4$. {Moreover, we have $\pmb{R}_{\text{AP},m}=\tilde{\pmb{R}}_{\text{AP},m}\tilde{\pmb{R}}_{\text{AP},m}^H$ with $\tilde{\pmb{R}}_{\text{AP},m}=d^{\text{PL}/2}[\pmb{A}\ \pmb{0}_{J\times J-P}]$, where $\pmb{A}=[\pmb{a}_{\phi_1},\ldots,\pmb{a}_{\phi_P}]\in\mathbb{C}^{J\times P}$. As a function of $\phi_p$, the $J$-dimensional steering vector is given by
\begin{align}
	\pmb{a}_{\phi_p}=\frac{1}{\sqrt{P}}\left[1,e^{-j2\pi \frac{d_{\text{AP}}}{\lambda}\sin(\phi_p)},\ldots,e^{-j2\pi \frac{d_{\text{AP}}}{\lambda}(J-1)\sin(\phi_p)}\right]^T,
\end{align}
where $\phi_p=-\pi/2+(p-1)\pi/P$ for $p=1,\ldots,P$. Furthermore, $d$ and $\text{PL}$ denote the communication distance and path loss corresponding to the scenario, respectively, and we utilize $P=J/2$ \cite{hoydis2013massive}. We leverage the Gaussian local scattering channel model to generate the spatial correlation matrix $\pmb{R}_{mu}$. Its $(p,q)$th element is given by \cite{10167480}
\begin{align}
	R_{mu}(p,q)=\frac{\beta_{mu}}{\sqrt{2\pi}\sigma_\chi}\int_{-\infty}^{\infty}e^{j2\pi d_H(p-q)\tilde{\chi}_{l_A,m}}e^{-\frac{\chi^2}{2\sigma^2_\chi}}d{\chi},
\end{align}
where $\tilde{\chi}_{l_A,m}=\sin(\theta_{l_A,m}+\chi)$, and $\chi\sim\mathcal{N}(0,\sigma^2_{\chi})$ with an angular standard deviation (ASD) of $\sigma_\chi$, and we use $\sigma_\chi=15^\circ$.} Unless stated otherwise, $M=20$ APs having $J=4$ antennas and $U=5$ UEs are employed. The RIS and AP element spacings are set as $d_\text{RIS}=\lambda/4$ and $d_\text{AP}=\lambda/2$, respectively. Each RIS includes $K=1$, $2$ and $4$ active RIS blocks, {it should be noted that the $K=G=4$ scenario is equivalent to the RIS-assisted CF-mMIMO system.} In the following figures, the conventional CF-mMIMO and RPM-RIS-assisted CF-mMIMO setups are denoted as ``Cell Free'' and ``RPM-RIS-CF'', respectively.
\begin{figure}[htbp]
\centering
\begin{minipage}[htbp]{\linewidth}
\centering
\includegraphics[width=0.9\linewidth]{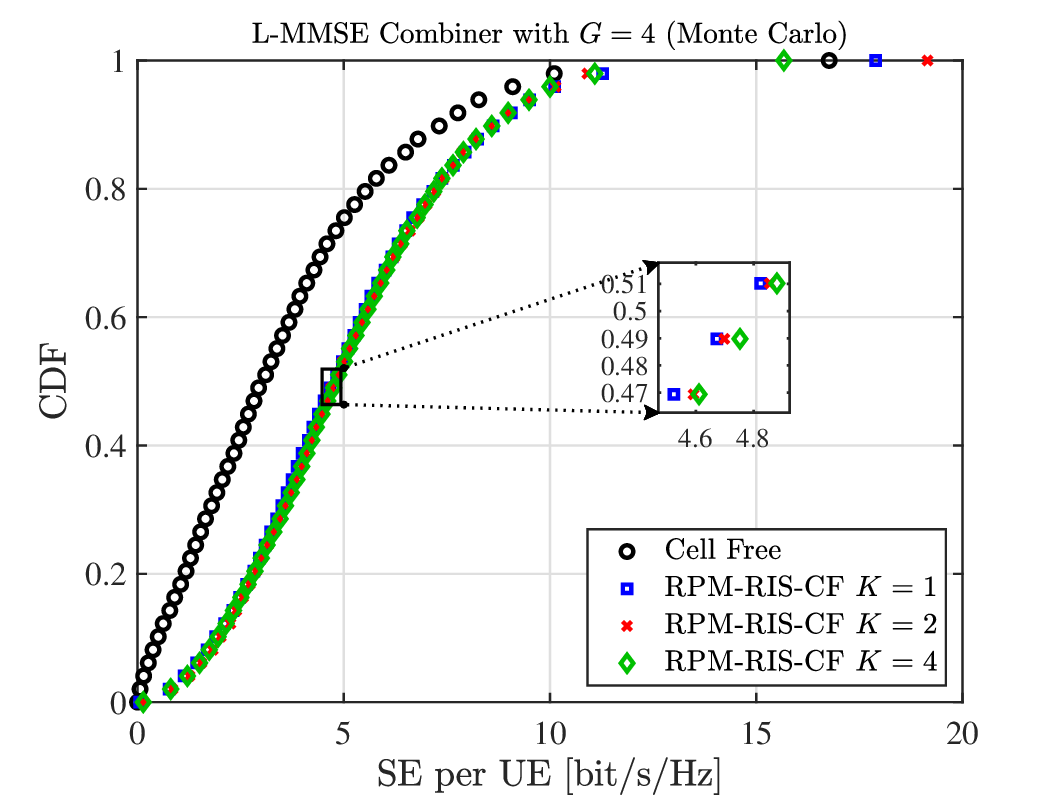}
\caption{CDF of SE for the CF-mMIMO and RPM-RIS-assisted CF-mMIMO systems with L-MMSE combiner ($G=4$ and $K=1, 2$ and $4$).}
\label{Figure2}
\end{minipage}
\hspace{0.01in}
\begin{minipage}[htbp]{\linewidth}
\centering
\includegraphics[width=0.9\linewidth]{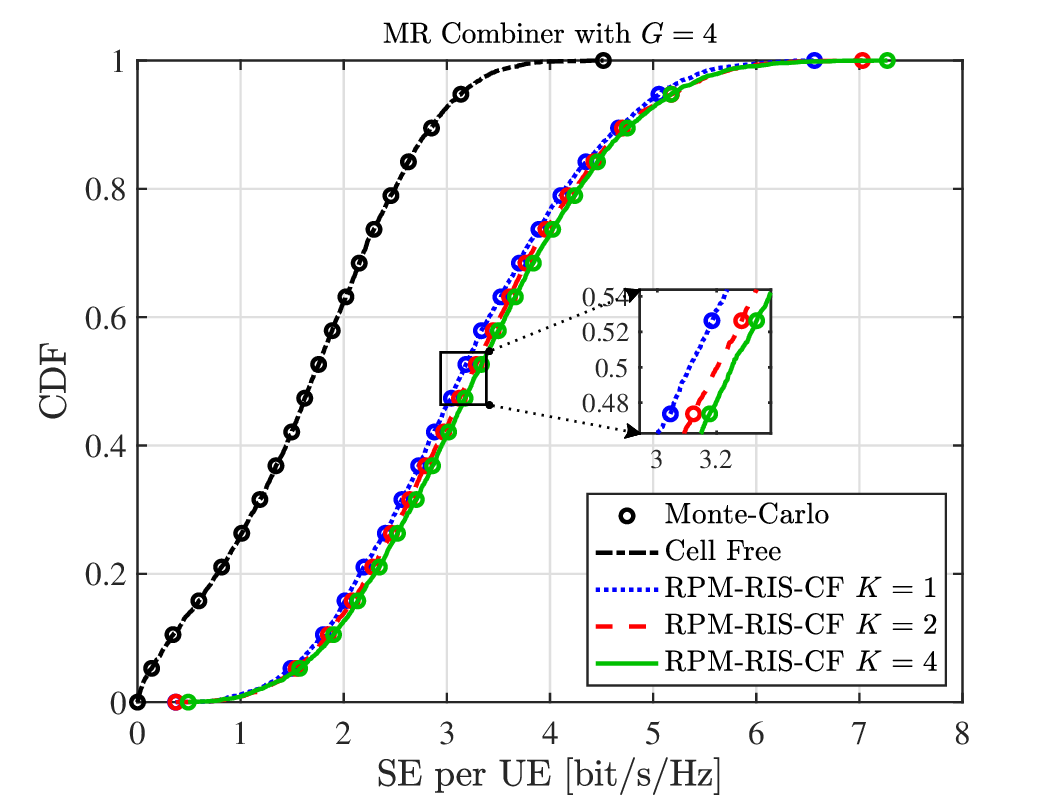}
\caption{CDF of SE for the CF-mMIMO and RPM-RIS-assisted CF-mMIMO systems with MR combiner ($G=4$ and $K=1, 2$ and $4$).}
\label{Figure3}
\end{minipage}
\end{figure}

In Fig. \ref{Figure2}, we investigate the cumulative distribution function (CDF) of the SE per UE associated with the conventional CF-mMIMO and our proposed RPM-RIS-assisted CF-mMIMO systems using Monte Carlo simulations based on \eqref{eq: SEu} and \eqref{eq: SINR monte carlo}, where the L-MMSE combiner and different numbers of active RIS blocks are considered. It can be observed that the RPM-RIS-assisted CF-mMIMO system significantly outperforms CF-mMIMO in terms of SE, which is consistent with \cite{van2021reconfigurable,10167480}. Moreover, RPM introduces a slight SE loss, as shown in Fig. \ref{Figure2}, since the cascaded channel gain is lower for less active RIS blocks. However, it should be emphasized that the CDF associated with $K=1$, $2$ and $4$ are nearly identical.

The CDFs of SE per UE for the CF-mMIMO and RPM-RIS-assisted CF-mMIMO systems are depicted in Fig. \ref{Figure3}. Moreover, the results of Monte Carlo simulations based on \eqref{eq: SEu} as well as \eqref{eq: SINR monte carlo} and the SE closed-form results based on \eqref{eq: SEu} and \eqref{eq: SINR closed form} are also illustrated. From Fig. \ref{Figure3}, we have the following observations. Firstly, a perfect overlap between the SE closed-form expression and the Monte Carlo simulations can be observed, which validates the accuracy of our analytical results. Secondly, the SE performance gaps of RPM-RIM-assisted CF-mMIMO associated with $K=1$ and $2$ compared to the RIS-assisted CF-mMIMO ($K=4$) are very small. Thirdly, as seen in Fig. \ref{Figure2} and Fig. \ref{Figure3}, the L-MMSE combiner is capable of attaining higher SE compared to the MR combiner, since the L-MMSE combiner can minimize the MSE at the cost of increased computational complexity, which is similar to conventional CF-mMIMO systems \cite{bjornson2019making}. Since the Monte Carlo simulation results are challenging to obtain, the closed-form SE derived based on the MR combiner will be utilized in the following figures.
\begin{figure}[t]
\vspace{-1em}
\centering
\begin{minipage}[htbp]{\linewidth}
\centering
\includegraphics[width=0.9\linewidth]{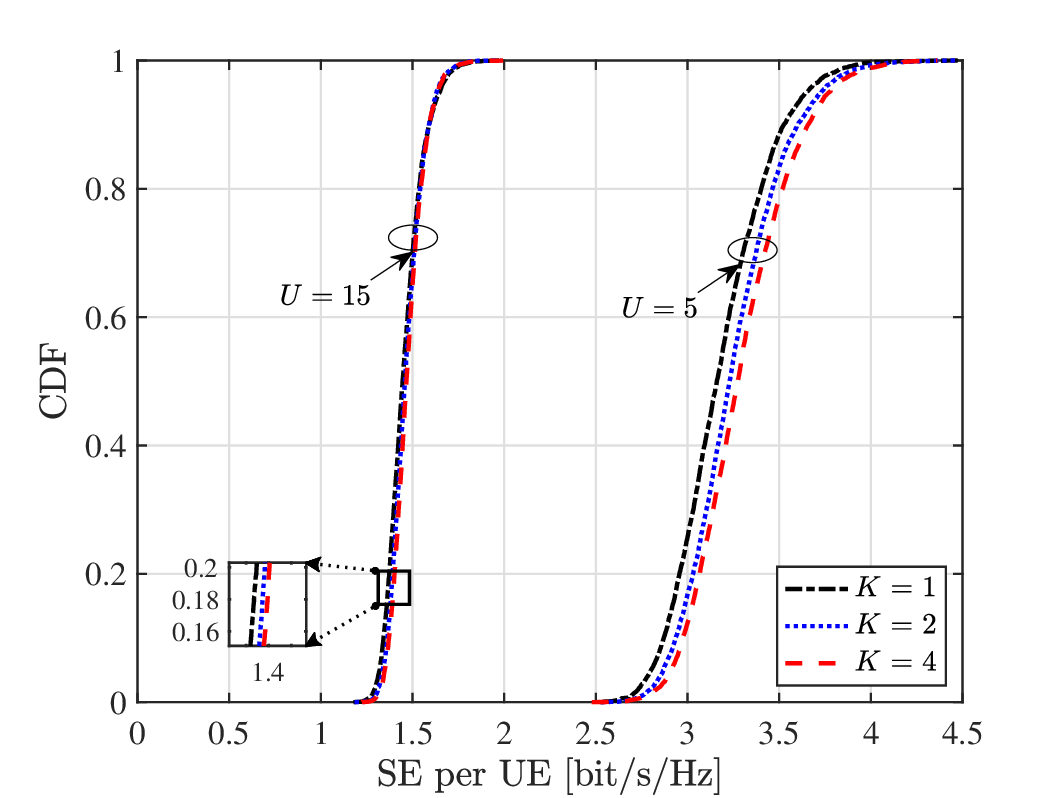}
\caption{CDF of SE per UE of the RPM-RIS-assisted CF-mMIMO systems with $U=[5,15]$ and $K=1, 2$ and $4$.}
\label{Figure4}
\end{minipage}
\hspace{0.01in}
\begin{minipage}[htbp]{\linewidth}
\centering
\includegraphics[width=0.9\linewidth]{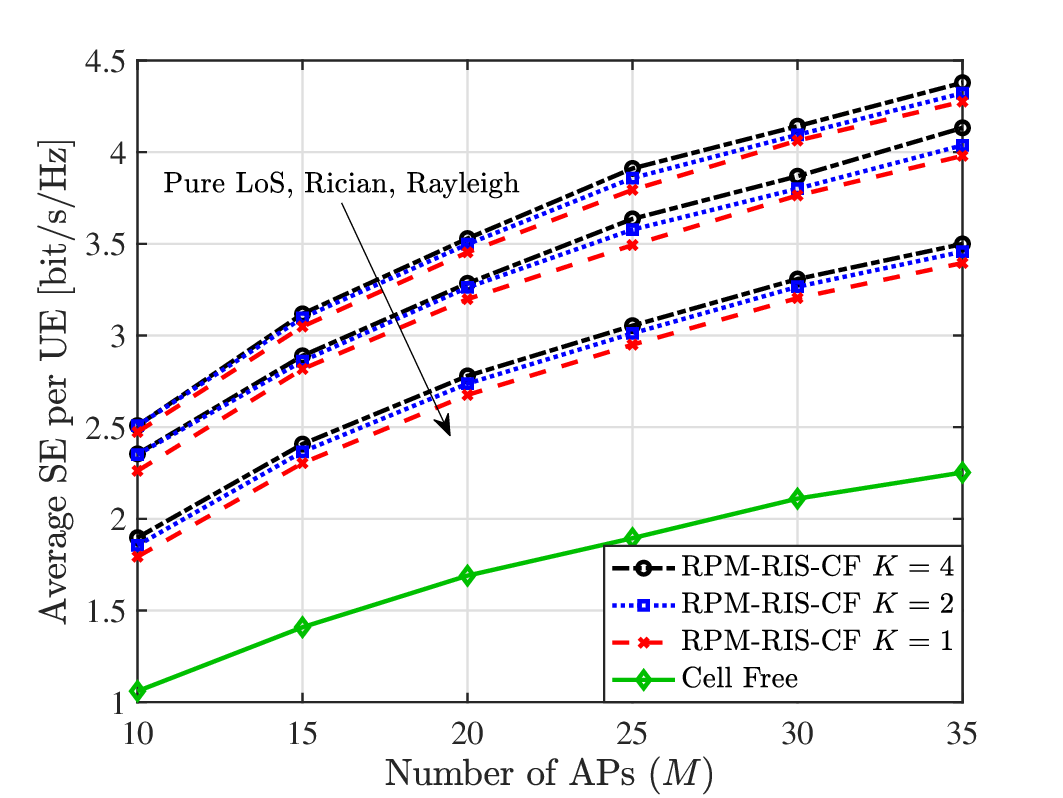}
\caption{{Average SE of CF-mMIMO and RPM-RIS-assisted CF-mMIMO systems with $G=4$ using different numbers of activated RIS blocks and different fading scenarios.}}
\label{Figure5}
\end{minipage}
\end{figure}

In Fig. \ref{Figure4}, the CDF of the SE per UE for the proposed RPM-RIM-assisted CF-mMIMO is investigated along with $K=1$, $2$ and $4$ active RIS blocks, while supporting $U=[5,15]$ UEs. It can be observed that higher SE can be achieved when employing $U=5$ UEs compared to the $U=15$ case. This is because lower pilot contamination and interference are achieved when fewer UEs are deployed and the pilot length $\tau_p$ is fixed. Furthermore, the SE gap introduced by RPM remains narrow, regardless of the value of $U$.

Figure \ref{Figure5} shows the average SE per UE of the CF-mMIMO and RPM-RIM-assisted CF-mMIMO systems as a function of the number of APs $M$ for different numbers of active RIS blocks $K$ and {different fading scenarios.} As shown in Fig. \ref{Figure5}, the SE can be enhanced by deploying more APs, regardless of the specific fading scenarios. This observation implies that both the number of spatial degrees of freedom and the channel hardening effect can be enhanced by increasing the value of $M$, yielding improved beamforming gain. Moreover, our proposed RPM-RIS-assisted CF-mMIMO scheme operating in Rician fading channels is capable of achieving up to $112.5\%$ higher SE than the conventional CF-mMIMO. {Furthermore, it can be observed that given the number of APs, Rician fading is capable of striking a higher average SE than in Rayleigh fading scenarios, while the pure LoS scenario exhibits the highest average SE. This is because the channel attenuation of Rician fading channels is statistically lower than the Rayleigh fading channels. Specifically, at $K=4$ and $M=35$, the Rician fading and pure LoS channels can attain about $18.1\%$ and $25.1\%$ SE improvements compared to that of Rayleigh fading channels, respectively.}\footnote{{We emphasize that the philosophy of our proposed architecture can be readily extended to other fading channel models, such as Nakagami-$m$, $\kappa-\mu$, etc \cite{zhang2015multivariate}.}}
\begin{figure}[htbp]
\vspace{-1em}
\centering
\begin{minipage}[htbp]{\linewidth}
\centering
\includegraphics[width=0.9\linewidth]{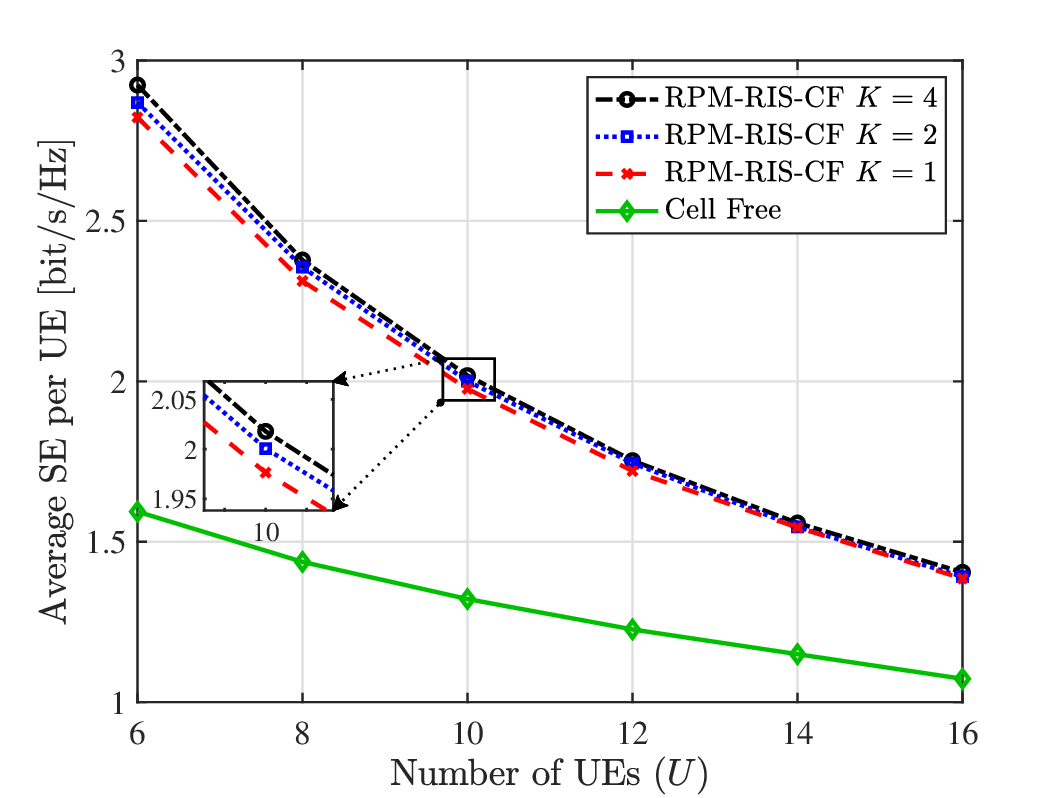}
\caption{Average SE of CF-mMIMO and RPM-RIS-assisted CF-mMIMO systems for $K=1, 2$ and $4$.}
\label{Figure6}
\end{minipage}
\hspace{0.01in}
\begin{minipage}[htbp]{\linewidth}
\centering
\includegraphics[width=0.9\linewidth]{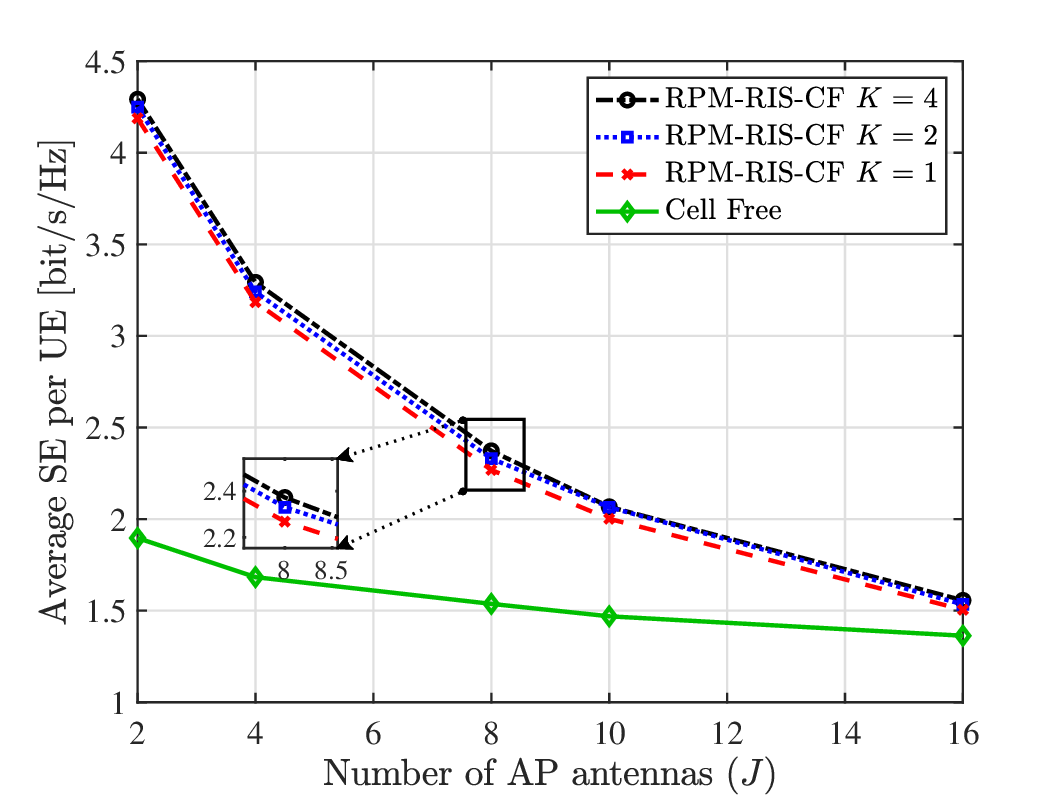}
\caption{Average SE of CF-mMIMO and RPM-RIS-assisted CF-mMIMO systems with $MJ=80$, for different numbers of activated RIS blocks.}
\label{Figure7}
\end{minipage}
\end{figure}

In Fig. \ref{Figure6}, the average SE of both the CF-mMIMO and RPM-RIS-assisted CF-mMIMO systems versus the number of UEs $U$ is evaluated. It can be observed that the average SE is reduced as the number of UEs increases. Moreover, the performance loss between $K=1$ and $2$ compared to the $K=4$ case becomes lower under the scenario of higher $U$ values, which is consistent with the findings in Fig. \ref{Figure4}. This observation can be explained by the fact that the interference among UEs is expected to be higher when more UEs are supported. Furthermore, given the $U=10$ case, our RPM-RIS-assisted CF-mMIMO can attain about $53.9\%$ SE gain compared to the conventional CF-mMIMO for any specific $K$.

In Fig. \ref{Figure7}, the average SE of both the CF-mMIMO and RPM-RIS-assisted CF-mMIMO systems are investigated as a function of the number of antennas per AP, while the total number of antennas is fixed as $MJ=80$. It can be seen that although the proposed RPM-RIS-assisted CF-mMIMO system still outperforms CF-mMIMO in terms of its average SE per UE, there is a significant performance erosion as $J$ increases. This trend can be explained by the following arguments. Firstly, since the value of $MJ$ remains unchanged, a higher $J$ implies that fewer APs are deployed, yielding higher path losses. Furthermore, lower macro-diversity can be obtained for a higher value of $J$. In addition, compared to the array gain, it can be readily shown that macro-diversity is the dominant factor for this scenario.

Next, the EE of our proposed RPM-RIS-assisted CF-mMIMO system is evaluated. The traffic-dependent power coefficients are set as $\varrho_m^{\text{AP}}=\varrho_m^{\text{BH}}=0.25$ (W/Gbps), $\forall m$. The fixed AP power consumption and the traffic-independent power consumed by each antenna are selected as $P^{\text{AP,fix}}_m=6$ W and $P_{m}^{\text{AP,a}}=0.15$ W, $\forall m$, respectively. The power amplifier efficiency is $\alpha^{\text{UE}}=0.4$. The fixed power consumed by the backhaul links is $P^{\text{BH,fix}}_m=0.8$ W, $\forall m$. Furthermore, the fixed power associated with the UEs and RIS elements is chosen as $P^{\text{UE,fix}}_u=10$ dBm and $P(b)=10/25$ dBm, $\forall u$, respectively.
\begin{figure}[htbp]
\vspace{-1em}
\centering
\begin{minipage}[htbp]{\linewidth}
\centering
\includegraphics[width=0.9\linewidth]{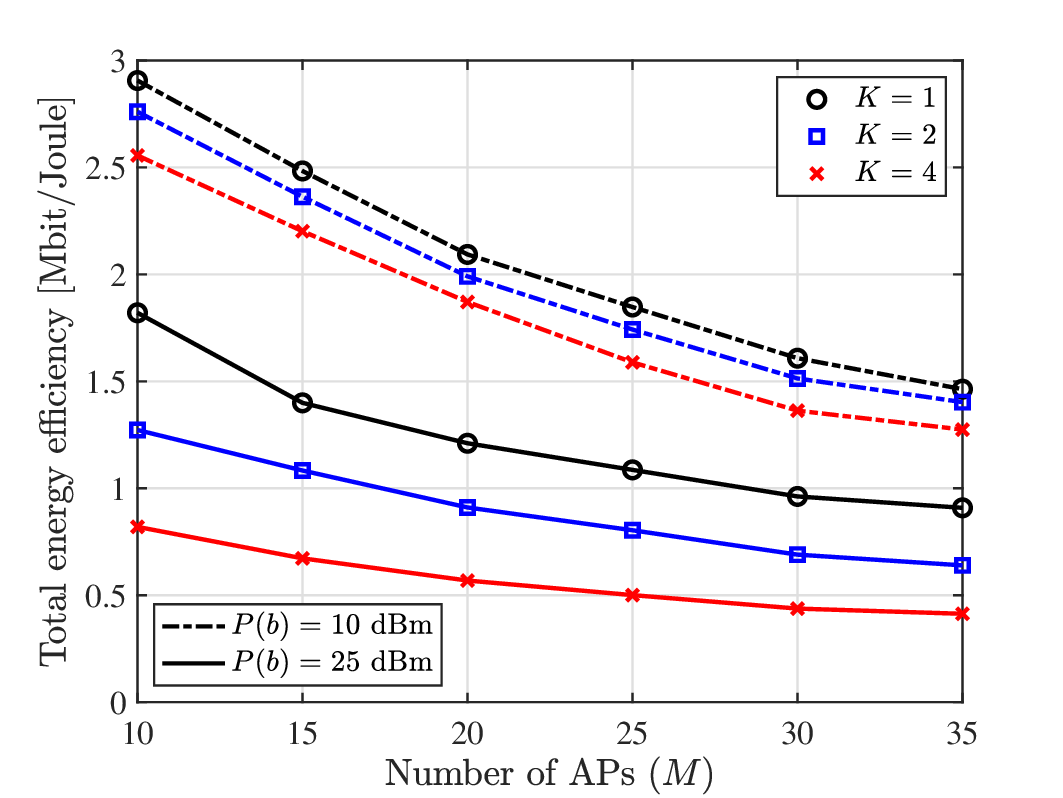}
\caption{Total EE of the RPM-RIS-assisted CF-mMIMO systems  with $K=1$, $2$, and $4$, when $P(b)=10$ dBm and $P(b)=25$ dBm.}
\label{Figure8}
\end{minipage}
\hspace{0.01in}
\begin{minipage}[htbp]{\linewidth}
\centering
\includegraphics[width=0.9\linewidth]{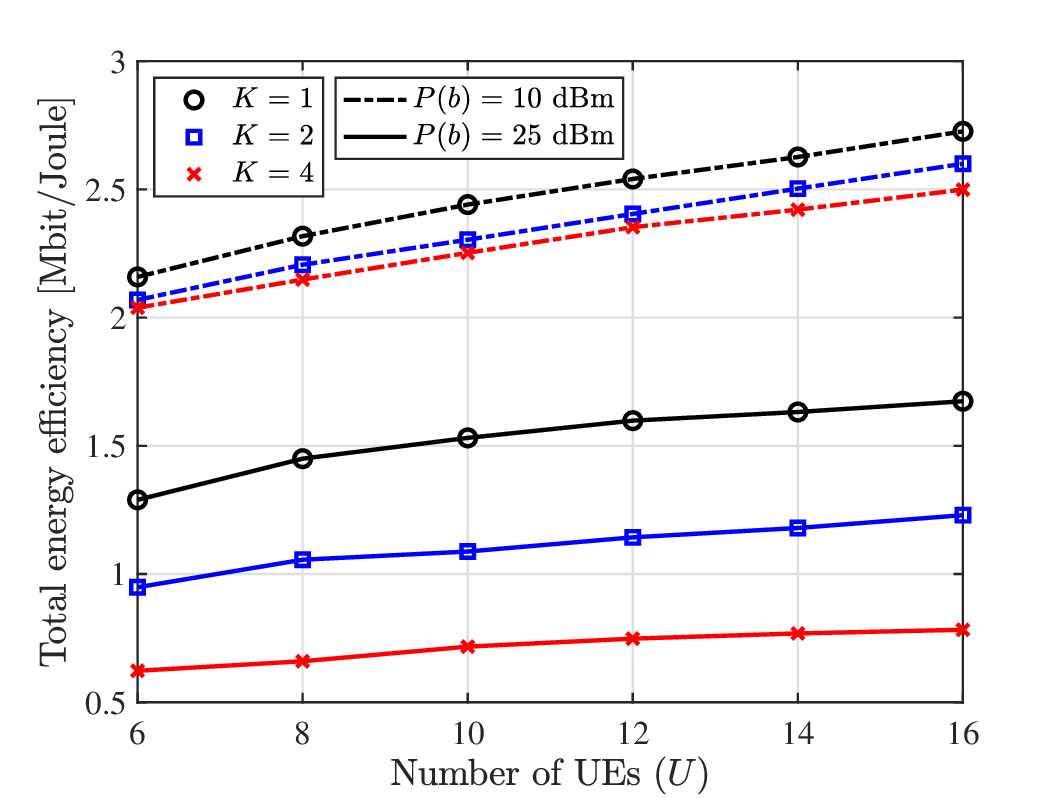}
\caption{Total EE of the proposed RPM-RIS-assisted CF-mMIMO systems with $K=1$, $2$, and $4$, when $P(b)=10$ dBm and $P(b)=25$ dBm.}
\label{Figure9}
\end{minipage}
\end{figure}

We compare the total EE of our proposed RPM-RIS-assisted CF-mMIMO system relying on random phase shifts, $K=[1,2,4]$ and $P(b)=[10,25]$ dBm in Fig. \ref{Figure8}, when different numbers of APs are employed. From Fig. \ref{Figure8}, we can make the following observations. Firstly, given a value of $K$, higher EE can be attained by selecting $P(b)=10$ dBm compared to the $P(b)=25$ dBm case. This is because the power consumed by the RISs is higher when using $P(b)=25$ dBm. Secondly, given a value of $P(b)$, the settings of using less active RIS blocks, i.e., $K=1$ results in better EE performance. This is because the total energy dissipation can be reduced by activating fewer RIS blocks, and the SE of the $K=1$ and $K=2$ cases is similar to the $K=4$ case, as shown in the figures of the SE part. In particular, compared to the RIS-assisted CF-mMIMO $(K=4$) system having $M=25$, the proposed RPM-RIS-assisted CF-mMIMO system with $K=1$ can improve the EE by $22.6\%$ and $120\%$ associated with $P(b)=10$ dBm and $P(b)=25$ dBm, respectively. Finally, an EE reduction can be observed as more APs are deployed, since the power consumption level increases and dominates the EE.

In Fig. \ref{Figure9}, the total EE associated random phase shifts is shown versus the number of UEs $U$ for different numbers of active RIS blocks and different values of $P(b)$. It can be observed that a lower $P(b)$ can significantly enhance the EE. Specifically, given $U=16$ and $K=2$, the setup with $P(b)=10$ dBm is capable of obtaining about twice higher total EE compared to the $P(b)=10$ dBm scenario. This observation implies that we should reduce $P(b)$ as much as possible, while sustaining a satisfactory SE to obtain better EE. Furthermore, the more UEs are supported, the better the total EE becomes. This is because the total SE can be enhanced while using more UEs, while the power dissipated by the UEs is relatively low compared to the other energy components of $P^{\text{tot}}$. Finally, according to Fig. \ref{Figure2} - Fig. \ref{Figure9}, it can be observed that our proposed RPM-RIS-assisted CF-mMIMO system is capable of striking an attractive SE \emph{vs.} EE trade-off.
\begin{figure}[htbp]
\centering
\vspace{-1em}
\subfigure[]{\label{Figure10-1}\includegraphics[width=0.9\linewidth]{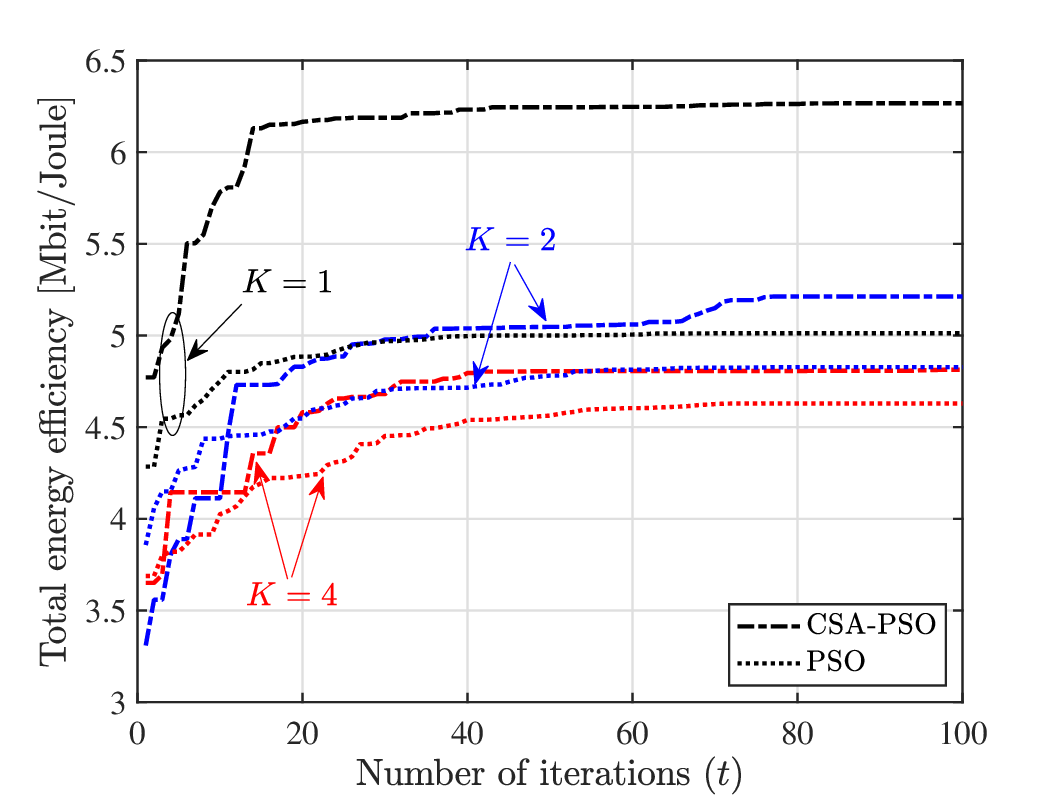}}
\subfigure[]{\label{Figure10-2}\includegraphics[width=0.9\linewidth]{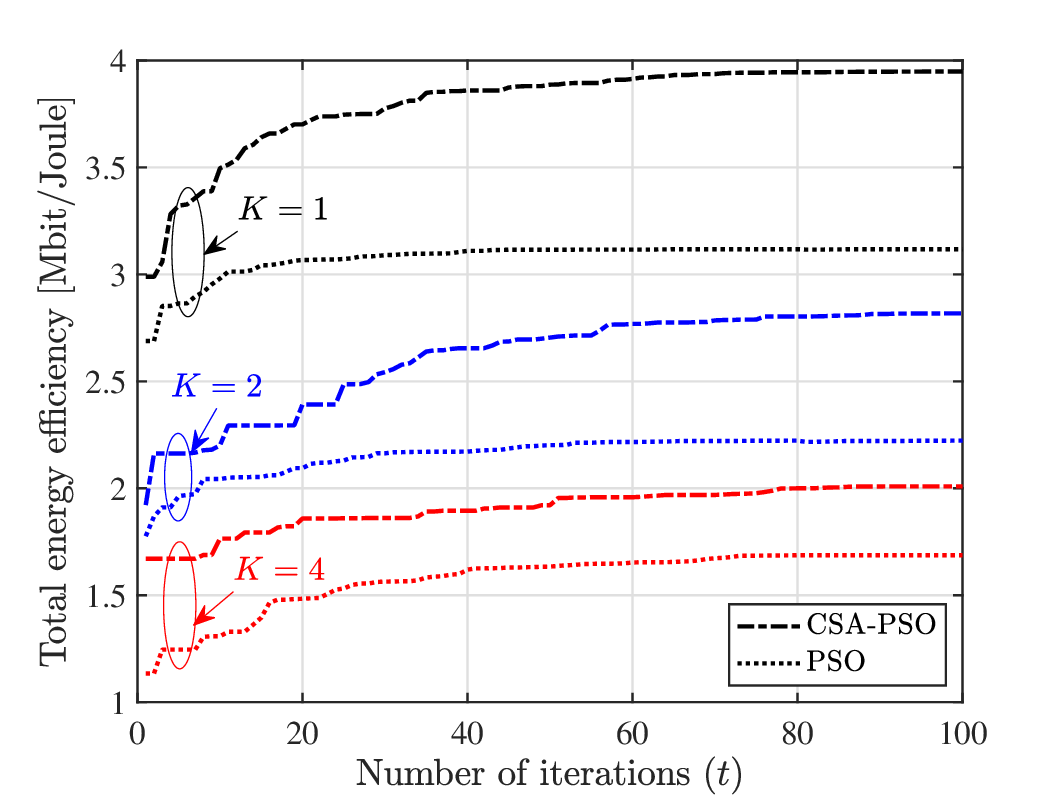}}
\caption{Total EE of the conventional PSO and the proposed CSA-PSO RIS phase shift design schemes operating at \subref{Figure10-1} $P(b)=10$ dBm and \subref{Figure10-2} $P(b)=25$ dBm, while $K=1$, $2$ and $4$.}
\label{Figure10}
\end{figure}

In Fig. \ref{Figure10}, we characterize the total EE of both our proposed CSA-PSO algorithm and of the conventional PSO algorithm. In particular, $M=8$ APs and $U=4$ UEs are considered. The maximum number of iterations is $T_{\text{max}}=100$, the value range of the velocity is given by $[-4,4]$, and we set $\omega=0.7298$ \cite{985692} in the conventional PSO algorithm. For the proposed CSA-PSO, we invoke $T_{\text{check}}=2$, $\omega_\text{max}=0.9$, $\omega_\text{max}=0.4$ and $N_i=10$. {It is clear that both CSA-PSO and PSO can converge within $40$ iterations.} Moreover, It can be observed that our proposed CSA-PSO is capable of significantly improving the EE of conventional PSO. More explicitly, under the condition of $K=1$, our CSA-PSO can obtain about $25\%$ and $26.7\%$ EE gain over PSO, for $P(b)=10$ dBm and $P(b)=25$ dBm, respectively. This is because of the following reasons. Firstly, we use a chaotic sequence to improve the particle's initial diversity. Secondly, an adaptive inertial weight factor is proposed to enhance the algorithm's search capability. Finally, the introduction of particle mutation and reset steps can help the particles escape from local optima.

\begin{figure}[htbp]
\vspace{-1em}
\centering
\includegraphics[width=0.9\linewidth]{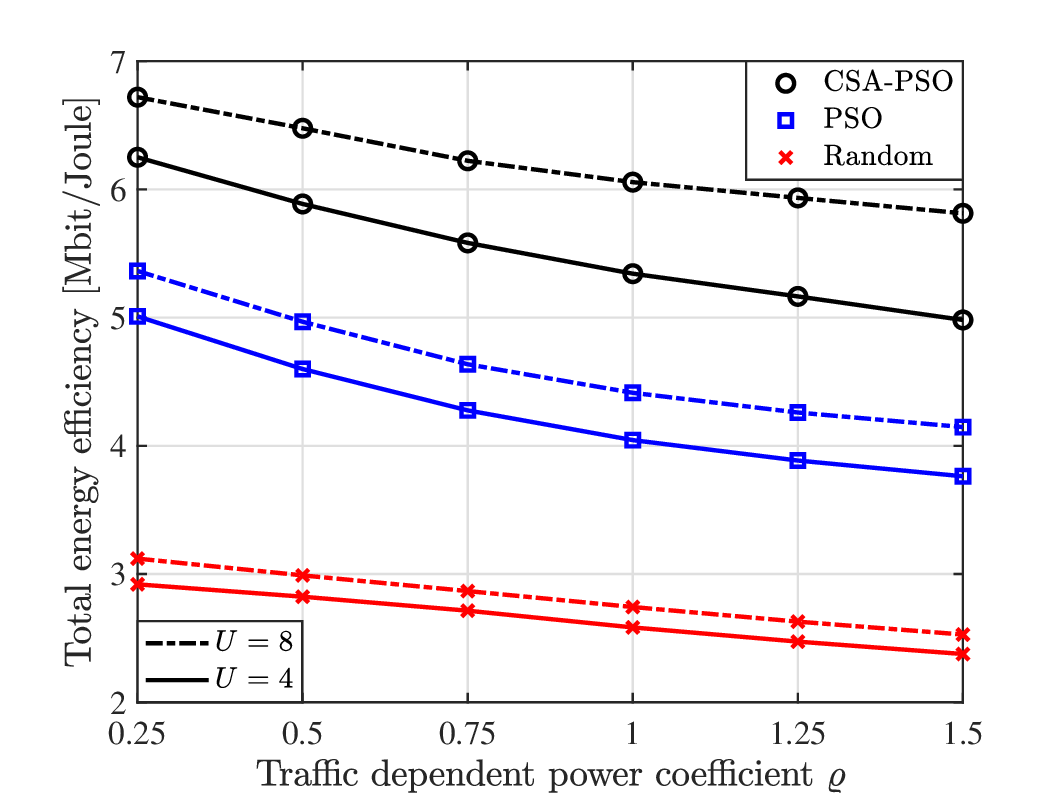}
\caption{Total EE of the proposed $K=2$ RPM-RIS-assisted CF-mMIMO systems with $P(b)=10$ dBm using $U=[4,8]$ and different RIS phase shift design schemes.}
\label{Figure11}
\vspace{-1em}
\end{figure}

The total EE are compared in Fig. \ref{Figure11}, as a function of the traffic-dependent power coefficient $\varrho_m^{\text{AP}}=\varrho_m^{\text{BH}}=\varrho$, $\forall m$ of our proposed CSA-PSO, and of the conventional PSO phase shift design and random phase shift schemes, where $U=[4,8]$ UEs are supported. From Fig. \ref{Figure11}, we have the following observations. First, compared to PSO and random phase shift, the proposed CSA-PSO algorithm is capable of attaining better EE performance. Specifically, given $U=4$ and $\varrho=1$, the CSA-PSO can achieve about $1.3\times$ and $2\times$ higher EE compared to its PSO and random phase shift counterparts. Secondly, the EE decreases as $\varrho$ increases, since higher energy is consumed. Thirdly, we can see that a higher EE can be obtained by supporting more UEs, consistent with our observations in Fig. \ref{Figure9}. Finally, the widest EE performance gap between $U=4$ and $U=8$ can be obtained by leveraging the CSA-PSO, followed by the conventional PSO and random phase shift schemes.
\begin{figure}[htbp]
\vspace{-1em}
\centering
\includegraphics[width=0.9\linewidth]{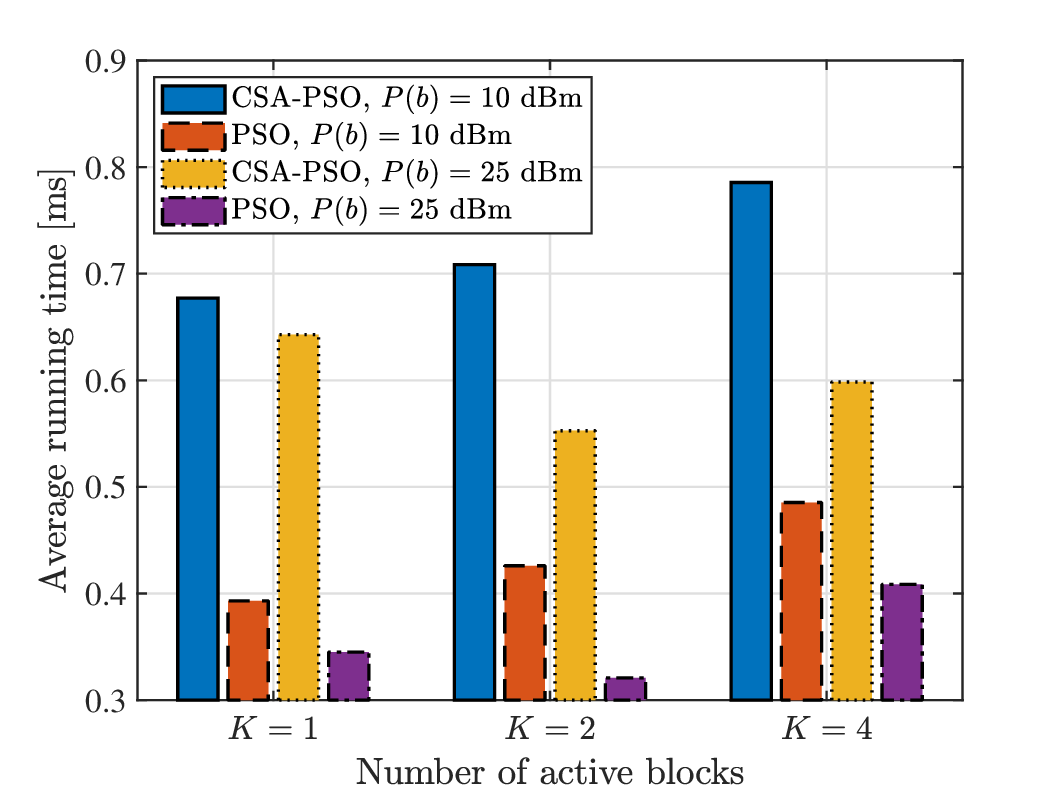}
\caption{{Average running time versus the number of active RIS blocks operating at $K=1,2,4$ using PSO and CSA-PSO-based phase shift optimization schemes with $P(b)=[10,25]$ dBm.}}
\label{Figure12}
\end{figure}

{In Fig. \ref{Figure12}, we compare the complexity of our proposed CSA-PSO algorithm and the conventional PSO in terms of the average running time under different numbers of active blocks with $P(b)=[10,25]$ dBm and the same parameters as in Fig. \ref{Figure10}. We leverage an \emph{Inter Core i9-14900HX} processor. It can be observed that given the values of $K$ and $P(b)$, PSO is more computationally efficient than CSA-PSO, which is consistent with our complexity analysis results in Section \ref{Section 3-3}. However, the CSA-PSO method can still converge within the coherence time of $T_c=1$ ms, regardless of the $K$ and $P(b)$ settings. Hence, based on Fig. \ref{Figure10}-Fig. \ref{Figure12}, we can readily state that our proposed CSA-PSO strikes an attractive complexity \emph{vs.} EE trade-off.}
\section{Summary and Conclusions}\label{Section 7}
In this paper, an RPM-RIS-assisted CF-mMIMO system was conceived, where only a part of RIS elements have been active, and extra information has been mapped onto the RIS ON/OFF states. Based on the uplink LSFD cooperation scheme, both the MR and L-MMSE combiners have been considered. Then, a closed-form expression was derived for the uplink SE by using the MR combiner, where both the channel estimation errors and pilot contamination were considered. Our simulation results demonstrated that the RPM-RIS-assisted CF-mMIMO system is capable of attaining nearly identical SE compared to the conventional RIS-assisted CF-mMIMO. Furthermore, the EE was characterized. For EE maximization, a CSA-PSO algorithm was proposed for RIS phase shift design. Moreover, the complexity of both CSA-PSO and classic PSO was analyzed. Our simulations illustrated that RPM-RIS-assisted CF-mMIMO attains a substantially better EE than RIS-assisted CF-mMIMO using a random phase shift. Finally, our proposed CSA-PSO scheme is capable of offering significantly higher EE compared to its conventional counterparts.

\begin{appendices}
	\section{Useful Lemmas}\label{appendix1}
	\emph{Lemma 1 {\cite[Lemma $7$]{9973349}:}} Assume that the elements of the matrix $\pmb{Z}\in\mathbb{C}^{m\times n}$ are i.i.d. RVs with zero mean and a variance of $\zeta_z$ with $m,n\geq 1$, and $\pmb{A}\in\mathbb{C}^{n\times n}$ is a deterministic matrix. Then, we have $\mathbb{E}\{\pmb{Z}\pmb{A}\pmb{Z}^H\}=\zeta_z\tr\{\pmb{A}\}\pmb{I}_m$.
	
\emph{Lemma 2 {\cite[Lemma $4$]{van2021reconfigurable}:}} We consider an $M\times 1$ RV vector $\pmb{a}\sim\mathcal{CN}(\pmb{0},\pmb{R}_a)$ with $\pmb{R}_a\in\mathbb{C}^{M\times M}$, and a deterministic matrix $\pmb{W}\in\mathbb{C}^{M\times M}$. Then, we have $\mathbb{E}\left\{|\pmb{a}^H\pmb{W}\pmb{a}|^2\right\}=|\tr(\pmb{R}_a\pmb{W})|^2+\tr(\pmb{R}_a\pmb{W}\pmb{R}_a\pmb{W}^H)$.

\section{}\label{appendix2}
Based on \eqref{eq: channel}, it can be readily shown that
\begin{align}\label{eq: ap-b-1}
	\pmb{R}_{mu}^h=\mathbb{E}\left\{\tilde{\pmb{R}}_{mu,1}^h+\tilde{\pmb{R}}_{mu,2}^h+\tilde{\pmb{R}}_{mu,3}^h+\tilde{\pmb{R}}_{mu,4}^h\right\},
\end{align}
where we have $\tilde{\pmb{R}}_{mu,1}^h=b_{mu}\kappa_m\bar{\pmb{G}}_m\pmb{\Phi}_m\tilde{\pmb{z}}_{mu}\tilde{\pmb{z}}_{mu}^H\pmb{\Phi}_m^H\bar{\pmb{G}}_m^H$, $\tilde{\pmb{R}}_{mu,2}^h=b_{mu}\iota_{mu}\tilde{\pmb{G}}_m\pmb{\Phi}_m\bar{\pmb{z}}_{mu}\bar{\pmb{z}}_{mu}^H\pmb{\Phi}_m^H\tilde{\pmb{G}}_m^H$, $\tilde{\pmb{R}}_{mu,3}^h=b_{mu}\tilde{\pmb{G}}_m\pmb{\Phi}_m\tilde{\pmb{z}}_{mu}\tilde{\pmb{z}}_{mu}^H\pmb{\Phi}_m^H\tilde{\pmb{G}}_m^H$ and $\tilde{\pmb{R}}_{mu,4}^h=\pmb{f}_{mu}\pmb{f}_{mu}^H$. Then, we have $\mathbb{E}\left\{\tilde{\pmb{R}}_{mu,1}^h\right\}=b_{mu}\kappa_m\bar{\pmb{G}}_m\pmb{\Phi}_m\tilde{\pmb{R}}_{mu}\pmb{\Phi}_m^H\bar{\pmb{G}}_m^H$ and $\mathbb{E}\left\{\tilde{\pmb{R}}_{mu,4}^h\right\}=\pmb{R}_{mu}$. Moreover, we have
\begin{align}\label{eq: ap-b-3}	
\pmb{\Pi}_{mu}&=\mathbb{E}\left\{\tilde{\pmb{R}}_{mu,2}^h\right\}=b_{mu}\iota_{mu}\mathbb{E}\left\{\tilde{\pmb{G}}_m\pmb{\Phi}_m\bar{\pmb{z}}_{mu}\bar{\pmb{z}}_{mu}^H\pmb{\Phi}^H_m\tilde{\pmb{G}}_m^H\right\}\nonumber\\
	&=b_{mu}\iota_{mu}\tilde{\pmb{\Pi}}_{mu}.
\end{align}
Let $\pmb{B}_{mu}\triangleq\pmb{\Phi}_m\bar{\pmb{z}}_{mu}\bar{\pmb{z}}_{mu}^H\pmb{\Phi}^H_m$, then the $(j,j')$th component of $\tilde{\pmb{\Pi}}_{mu}$, i.e., $\mathbb{E}\left\{\tilde{\pmb{G}}_m(j,:)\pmb{B}_{mu}\tilde{\pmb{G}}_m(j',:)^H\right\}$ can be formulated as $\tr\left(\mathbb{E}\left\{\pmb{B}_{mu}\tilde{\pmb{G}}_m(j',:)^H\tilde{\pmb{G}}_m(j,:)\right\}\right)=\tr\left(\pmb{B}_{mu}\tilde{\pmb{R}}_m(jL_A-L+1:jL_A,j'L_A-L_A+1:j'L_A)\right)$. Moreover, we have $\pmb{\Xi}_{mu}=\mathbb{E}\left\{\tilde{\pmb{R}}_{mu,3}^h\right\}=b_{mu}\mathbb{E}\left\{\tilde{\pmb{G}}_m\pmb{\Phi}_m\tilde{\pmb{R}}_{mu}\pmb{\Phi}^H_m\tilde{\pmb{G}}_m^H\right\}$. Let $\tilde{\pmb{B}}_{mu}\triangleq\pmb{\Phi}_m\tilde{\pmb{R}}_{mu}\pmb{\Phi}^H_m$, then the $(j,j')$th element of $\pmb{\Xi}_{mu}$ can be expressed as $\tr\left(\tilde{\pmb{B}}_{mu}\tilde{\pmb{R}}_m(jL_A-L_A+1:jL_A,j'L_A-L_A+1:j'L_A)\right)$. Consequently, the proof can be completed by combining the above results.
\vspace{-2em}
	\section{}\label{appendix3}
We first offer several useful results that are utilized in the proof. Given the set $\mathcal{P}_u$ and $k\in\mathcal{P}_u\setminus\{u\}$, we have
\begin{align}\label{eq: ap-c-useful2}
	\pmb{y}^p_{mu}=\pmb{y}^p_{mk},\ \bar{\pmb{y}}^p_{mu}=\bar{\pmb{y}}^p_{mk},\ \pmb{\Psi}_{mu}=\pmb{\Psi}_{mk}.
\end{align}
Based on \emph{Proposition 1}, it can be observed that $\pmb{R}_{mu}^h$ are Hermitian matrices. Hence, we have $\pmb{R}_{mu}^h=(\pmb{R}_{mu}^h)^H$ and $\pmb{\Psi}_{mu}^h=(\pmb{\Psi}_{mu}^h)^H, \forall m,u$. Furthermore, it can be readily shown based on \eqref{eq: channel estimate} that
\begin{align}\label{eq: ap-c-1}
	\pmb{y}^p_{mu}-\bar{\pmb{y}}^p_{mu}\sim\mathcal{CN}(\pmb{0},{\tau_p}\pmb{\Psi}_{mu}).
\end{align}

Now we calculate each term of \eqref{eq: SINR monte carlo}. Since the channel estimate $\hat{\pmb{h}}_{mu}$ and the corresponding error $\tilde{\pmb{h}}_{mu}$ are uncorrelated, for the numerator term we have $\mathbb{E}\{\pmb{g}_{uu}\}=\mathbb{E}\left\{\left[\hat{\pmb{h}}_{1u}^H\hat{\pmb{h}}_{1u},\ldots,\hat{\pmb{h}}_{Mu}^H\hat{\pmb{h}}_{Mu}\right]^T\right\}$. Consequently, let $\tilde{\pmb{y}}^p_{mu}=\pmb{y}^p_{mu}-\bar{\pmb{y}}^p_{mu}$, hence the $m$-th term of $\mathbb{E}\{\pmb{g}_{uu}\}$, i.e., $\mathbb{E}\left\{\hat{\pmb{h}}_{mu}^H\hat{\pmb{h}}_{mu}\right\}$ can be derived based on \eqref{eq: channel estimate} as
\begin{align}\label{eq: ap-c-2}
&\left\|\bar{\pmb{h}}_{mu}\right\|^2+p_u\tr\left(\mathbb{E}\{(\tilde{\pmb{y}}^p_{mu})^H(\pmb{\Psi}_{mu}^{-1})(\pmb{R}_{mu}^h)\pmb{R}_{mu}^h\pmb{\Psi}_{mu}^{-1}(\tilde{\pmb{y}}^p_{mu})\}\right)\nonumber\\
	&\overset{(a)}{=}\left\|\bar{\pmb{h}}_{mu}\right\|^2+p_u\tr\left(\mathbb{E}\{\pmb{R}_{mu}^h\pmb{\Psi}_{mu}^{-1}(\tilde{\pmb{y}}^p_{mu})(\tilde{\pmb{y}}^p_{mu})^H\pmb{\Psi}_{mu}^{-1}\pmb{R}_{mu}^h\}\right)\nonumber\\
	&\overset{(b)}{=}\left\|\bar{\pmb{h}}_{mu}\right\|^2+p_u\tau_p\tr(\pmb{R}_{mu}^h\pmb{\Psi}_{mu}^{-1}\pmb{\Psi}_{mu}\pmb{\Psi}_{mu}^{-1}\pmb{R}_{mu}^h)\nonumber\\
	&=\left\|\bar{\pmb{h}}_{mu}\right\|^2+p_u\tau_p\tr(\pmb{\Gamma}_{mu})=\bar{\xi}_{mu},
	\end{align}
where $(a)$ invokes $\tr(\pmb{X}\pmb{Y})=\tr({\pmb{Y}\pmb{X}})$ along with the appropriate dimensions of matrices $\pmb{X}$ and $\pmb{Y}$, and $(b)$ can be obtained by virtue of \eqref{eq: ap-c-1}. Hence, we have $\mathbb{E}\{\pmb{g}_{uu}\}=[\bar{\xi}_{1u},\ldots,\bar{\xi}_{Mu}]^T$ and the numerator term can be readily obtained.

Similarly, we consider the noise term of \eqref{eq: SINR monte carlo}:
\begin{align}\label{eq: ap-c-3}
	\pmb{c}^H_u\pmb{V}_u\pmb{c}_u&=\pmb{c}^H_u\diag\left\{\mathbb{E}\left\{||\hat{\pmb{h}}_{1u}||\right\},\ldots,\mathbb{E}\left\{||\hat{\pmb{h}}_{Mu}||\right\}\right\}\pmb{c}_u\nonumber\\
	&=\tr\left(\pmb{C}_u^H\diag\{\bar{\xi}_{1u},\ldots,\bar{\xi}_{Mu}\}\pmb{C}_u\right).
\end{align}
For the second term of the denominator $\pmb{c}^H_u\mathbb{E}\{\pmb{g}_{uu}\}\mathbb{E}\{\pmb{g}_{uu}^H\}\pmb{c}_u$, it can be formulated as
\begin{align}\label{eq: ap-c-4}	
&\left\{\sum_{m=1}^M c_{mu}\left[\tr(p_u\tau_p\pmb{\Gamma}_{mu})+\left\|\bar{\pmb{h}}_{mu}\right\|^2\right]\right\}\nonumber\\
&\times\left\{\sum_{n=1}^M c_{nu}^*\left[\tr(p_u\tau_p\pmb{\Gamma}_{nu})+\left\|\bar{\pmb{h}}_{nu}\right\|^2\right]\right\}.
		\end{align}
When $m=n$, the first component of \eqref{eq: ap-c-4} can be attained as
\begin{align}\label{eq: ap-c-5}	
&\sum_{m=1}^M |c_{mu}|^2\left[p_u^2\tau_p^2|\tr(\pmb{\Gamma}_{mu})|^2+\left\|\bar{\pmb{h}}_{mu}\right\|^2\left\|\bar{\pmb{h}}_{mu}\right\|^2\right.\nonumber\\
&+\left.2p_u\tau_p\tr(\pmb{\Gamma}_{mu})\left\|\bar{\pmb{h}}_{mu}\right\|^2\right].
\end{align}
In the case of $m\neq n$, the second component of \eqref{eq: ap-c-4} can be formulated as
\begin{align}\label{eq: ap-c-6}	
\sum_{m=1}^M\sum_{n=1}^M& c_{mu}c^*_{nu}\left[p_u^2\tau_p^2\tr\left(\pmb{\Gamma}_{mu}\right)\tr(\pmb{\Gamma}_{nu})+\tr\left(\bar{\pmb{h}}_{mu}^H\bar{\pmb{h}}_{mu}\right)\right.\nonumber\\
&\times\tr\left(\bar{\pmb{h}}_{nu}^H\bar{\pmb{h}}_{nu}\right)+p_u\tau_p\tr\left(\bar{\pmb{h}}_{mu}^H\bar{\pmb{h}}_{mu}\right)\tr\left(\pmb{\Gamma}_{nu}\right)\nonumber\\
	&\left.+p_u\tau_p\tr\left(\bar{\pmb{h}}_{nu}^H\bar{\pmb{h}}_{nu}\right)\tr\left(\pmb{\Gamma}_{mu}\right)\right].
	\end{align}
\end{appendices}Therefore, $\pmb{c}^H_u\mathbb{E}\{\pmb{g}_{uu}\}\mathbb{E}\{\pmb{g}_{uu}^H\}\pmb{c}_u$ is calculated by combining \eqref{eq: ap-c-5} and \eqref{eq: ap-c-6}. Furthermore, the interference term $\mathbb{E}\left\{\left|\sum_{m=1}^M c_{mu}\hat{\pmb{h}}^H_{mu}{\pmb{h}}_{mk}\right|^2\right\}$ can be expressed as $\sum_{m=1}^M\sum_{n=1}^M c_{mu}c^*_{nu}\mathbb{E}\left\{\left(\hat{\pmb{h}}_{mu}^H{\pmb{h}}_{mk}\right)^H\left(\hat{\pmb{h}}_{nu}^H{\pmb{h}}_{nk}\right)\right\}$, where all the combinations of the AP and UE indices have to be considered. Since the estimated channels of different APs are independent, we have $\mathbb{E}\left\{\left(\hat{\pmb{h}}_{mu}^H{\pmb{h}}_{mk}\right)^H\left(\hat{\pmb{h}}_{nu}^H{\pmb{h}}_{nk}\right)\right\}=0$ when $m\neq n$ and $k\notin\mathcal{P}_u$. Under the condition of $m\neq n$ and $k\in\mathcal{P}_u\setminus\{u\}$, we have $\mathbb{E}\left\{\left(\hat{\pmb{h}}_{mu}^H{\pmb{h}}_{mk}\right)^H\left(\hat{\pmb{h}}_{nu}^H{\pmb{h}}_{nk}\right)\right\}=\mathbb{E}\left\{\left(\hat{\pmb{h}}_{mu}^H{\hat{\pmb{h}}}_{mk}\right)^H\right\}\mathbb{E}\left\{\left(\hat{\pmb{h}}_{nu}^H{\hat{\pmb{h}}}_{nk}\right)\right\}$, and the component $\mathbb{E}\left\{\left(\hat{\pmb{h}}_{mu}^H{\hat{\pmb{h}}}_{mk}\right)^H\right\}$ can be formulated as
\begin{align}\label{eq: ap-c-8}	
	&\mathbb{E}\left\{\left[\bar{\pmb{h}}_{mu}^H+\sqrt{p_u}(\tilde{\pmb{y}}^p_{mu})^H(\pmb{\Psi}_{mu}^{-1})^H(\pmb{R}_{mu}^h)^H\right]\right.\nonumber\\
	&\times\left.\left[\bar{\pmb{h}}_{mk}+\sqrt{p_k}\pmb{R}_{mk}^h\pmb{\Psi}_{mk}^{-1}(\tilde{\pmb{y}}^p_{mk})\right]\right\}\nonumber\\
	&\overset{(a)}{=}\sqrt{\tilde{p}_{uk}}\mathbb{E}\left\{(\tilde{\pmb{y}}^p_{mk})^H(\pmb{\Psi}_{mu}^{-1})^H(\pmb{R}_{mu}^h)^H\pmb{R}_{mk}^h\pmb{\Psi}_{mk}^{-1}\tilde{\pmb{y}}^p_{mk}\right\}\nonumber\\
	&\overset{(b)}{=}\sqrt{\tilde{p}_{uk}}\tr\left(\mathbb{E}\left\{(\pmb{\Psi}_{mu}^{-1})^H(\pmb{R}_{mu}^h)^H\pmb{R}_{mk}^h\pmb{\Psi}_{mk}^{-1}\tilde{\pmb{y}}^p_{mk}(\tilde{\pmb{y}}^p_{mk})^H\right\}\right)\nonumber\\
	&=\sqrt{\tilde{p}_{uk}}\tau_p\tr(\pmb{R}_{mk}^h\pmb{\Psi}_{mu}^{-1}\pmb{R}_{mu}^h),
\end{align}
where $\tilde{p}_{uk}=p_u p_k$, $(a)$ invokes $\mathbb{E}\left\{\bar{\pmb{h}}_{mu}^H\bar{\pmb{h}}_{mk}\right\}=0$ and $\mathbb{E}\left\{(\tilde{\pmb{y}}_{mu}^p)^H(\pmb{\Psi}_{mu}^{-1})^H(\pmb{R}_{mu}^h)^H\bar{\pmb{h}}_{mk}\right\}=0$, $(b)$ uses \eqref{eq: ap-c-useful2} and $\tr(\pmb{X}\pmb{Y})=\tr({\pmb{Y}\pmb{X}})$. Hence, it can be readily shown that $\mathbb{E}\left\{\left(\hat{\pmb{h}}_{mu}^H{\hat{\pmb{h}}}_{mk}\right)^H\right\}\mathbb{E}\left\{\hat{\pmb{h}}_{nu}^H{\hat{\pmb{h}}}_{nk}\right\}=p_u p_k\tau_p^2\tr\left(\pmb{R}_{mk}^h\pmb{\Psi}_{mu}^{-1}\pmb{R}_{mu}^h\right)\tr\left(\pmb{R}_{nu}^h\pmb{\Psi}_{nu}^{-1}\pmb{R}_{nk}^h\right)$. When $m\neq n$ and $k=u$, we express $\mathbb{E}\left\{\left(\hat{\pmb{h}}_{mu}^H{\pmb{h}}_{mk}\right)^H\left(\hat{\pmb{h}}_{nu}^H{\pmb{h}}_{nk}\right)\right\}=\mathbb{E}\left\{\hat{\pmb{h}}_{mu}^H{\hat{\pmb{h}}}_{mk}\right\}\mathbb{E}\left\{\hat{\pmb{h}}_{nu}^H{\hat{\pmb{h}}}_{nk}\right\}$ since the channel estimates at different APs are independent, while the channel estimation error and the estimated channel are uncorrelated. In addition, we have $\mathbb{E}\left\{\hat{\pmb{h}}_{mu}^H{\hat{\pmb{h}}}_{mk}\right\}=\tr\left(\bar{\pmb{h}}_{mu}^H\bar{\pmb{h}}_{mu}\right)+p_u\tau_p\tr(\pmb{\Gamma}_{mu})$. Therefore, we have
\begin{align}\label{eq: ap-c-10}	
	&\mathbb{E}\left\{\hat{\pmb{h}}_{mu}^H{\hat{\pmb{h}}}_{mk}\right\}\mathbb{E}\left\{\hat{\pmb{h}}_{nu}^H{\hat{\pmb{h}}}_{nk}\right\}=\tr\left(\bar{\pmb{h}}_{mu}^H\bar{\pmb{h}}_{mu}\right)\tr\left(\bar{\pmb{h}}_{nu}^H\bar{\pmb{h}}_{nu}\right)\nonumber\\
	&+p_u^2\tau_p^2\tr(\pmb{\Gamma}_{mu})\tr(\pmb{\Gamma}_{nu})+p_k\tau_p\tr\left(\bar{\pmb{h}}_{mu}^H\bar{\pmb{h}}_{mu}\right)\tr(\pmb{\Gamma}_{nu})\nonumber\\&+p_k\tau_p\tr\left(\bar{\pmb{h}}_{nu}^H\bar{\pmb{h}}_{nu}\right)\tr(\pmb{\Gamma}_{mu}).
\end{align}
In the case of $m=n$ and $k=u$, the interference includes the following two terms $\mathbb{E}\left\{\left|\hat{\pmb{h}}_{mu}^H\tilde{\pmb{h}}_{mu}\right|^2\right\}=\tr\left(\left(p_u\tau_p\pmb{\Gamma}_{mu}+\bar{\pmb{h}}_{mu}^H\bar{\pmb{h}}_{mu}\right)\pmb{\Lambda}_{mu}\right)$ and $\mathbb{E}\left\{\left|\hat{\pmb{h}}_{mu}^H\hat{\pmb{h}}_{mu}\right|^2\right\}$ can be formulated as \eqref{eq: ap-c-12} at the top of next page, where $(a)$ is obtained by using \eqref{eq: ap-c-useful2} with $\pmb{w}\sim\mathcal{CN}(\pmb{0},\pmb{I}_J)$. By denoting $a=\bar{\pmb{h}}_{mu}^H\bar{\pmb{h}}_{mu}$, $b=\sqrt{p_u\tau_p}\pmb{w}^H(\pmb{\Psi}_{mu}^{-1/2})^H(\pmb{R}_{mu}^h)^H\bar{\pmb{h}}_{mu}$, $c=\sqrt{p_u\tau_p}\bar{\pmb{h}}_{mu}^H\pmb{R}_{mu}^h\pmb{\Psi}_{mu}^{-1/2}\pmb{w}$ and $d=\left|\sqrt{p_u\tau_p}\pmb{R}_{mu}^h\pmb{\Psi}_{mu}^{-1/2}\pmb{w}\right|^2$, \eqref{eq: ap-c-12} can be rewritten as $\mathbb{E}\left\{\left|\hat{\pmb{h}}_{mu}^H\hat{\pmb{h}}_{mu}\right|^2\right\}=\mathbb{E}\{|a|^2\}+\mathbb{E}\{|b|^2\}+\mathbb{E}\{|c|^2\}+\mathbb{E}\{|d|^2\}+2\mathbb{E}\{ad\}$.
\setcounter{eqnback}{\value{equation}} \setcounter{equation}{32}
\begin{figure*}[!t]
\begin{align}\label{eq: ap-c-12}
&\mathbb{E}\left\{\left|\hat{\pmb{h}}_{mu}^H\hat{\pmb{h}}_{mu}\right|^2\right\}=\mathbb{E}\left\{\left|\left[\bar{\pmb{h}}_{mu}+\sqrt{p_u}\pmb{R}^h_{mu}\pmb{\Psi}_{mu}^{-1}(\tilde{\pmb{y}}^p_{mu})\right]^H\left[\bar{\pmb{h}}_{mu}+\sqrt{p_u}\pmb{R}^h_{mu}\pmb{\Psi}_{mu}^{-1}(\tilde{\pmb{y}}^p_{mu})\right]\right|^2\right\}\nonumber\\
	&\overset{(a)}{=}\mathbb{E}\left\{\left|\left[\bar{\pmb{h}}_{mu}+\sqrt{p_u\tau_p}\pmb{R}^h_{mu}\pmb{\Psi}_{mu}^{-1/2}\pmb{w}\right]^H\left[\bar{\pmb{h}}_{mu}+\sqrt{p_u\tau_p}\pmb{R}^h_{mu}\pmb{\Psi}_{mu}^{-1/2}\pmb{w}\right]\right|^2\right\}\nonumber\\
	&=\mathbb{E}\left\{\left|\bar{\pmb{h}}_{mu}^H\bar{\pmb{h}}_{mu}\!+\!\sqrt{p_u\tau_p}\pmb{w}^H\pmb{\Psi}_{mu}^{-1/2}\pmb{R}_{mu}^h\bar{\pmb{h}}_{mu}\!+\!\sqrt{p_u\tau_p}\bar{\pmb{h}}_{mu}^H\pmb{R}_{mu}^h\pmb{\Psi}_{mu}^{-1/2}\pmb{w}+\left|\sqrt{p_u\tau_p}\pmb{R}_{mu}^h\pmb{\Psi}_{mu}^{-1/2}\pmb{w}\right|^2\right|^2\right\},
\end{align}
\hrulefill
\vspace{-2em}
\end{figure*}
\setcounter{eqncnt}{\value{equation}}
\setcounter{equation}{\value{eqnback}}
Consequently, we have $\mathbb{E}\{|a|^2\}=\bar{\pmb{h}}_{mu}^H\bar{\pmb{h}}_{mu}\bar{\pmb{h}}_{mu}^H\bar{\pmb{h}}_{mu}$. Then, we can also obtain $\mathbb{E}\{|b|^2\}=p_u\tau_p\mathbb{E}\left\{\pmb{w}^H(\pmb{\Psi}_{mu}^{-1/2})^H(\pmb{R}_{mu}^h)^H\bar{\pmb{h}}_{mu}\bar{\pmb{h}}_{mu}^H\pmb{R}_{mu}^h\pmb{\Psi}_{mu}^{-1/2}\pmb{w}\right\}\overset{(a)}{=}p_u\tau_p\bar{\pmb{h}}_{mu}^H\bar{\pmb{h}}_{mu}\tr\left(\pmb{\Gamma}_{mu}\right)$, where $(a)$ follows Lemma 1, the LoS components are determinstic and $\tr(\pmb{X}\pmb{Y})=\tr({\pmb{Y}\pmb{X}})$. Moreover, it can be readily shown that $\mathbb{E}\{|b|^2\}=\mathbb{E}\{|c|^2\}=\mathbb{E}\{ad\}=p_u\tau_p\bar{\pmb{h}}_{mu}^H\bar{\pmb{h}}_{mu}\tr\left(\pmb{\Gamma}_{mu}\right)$. Next, $\mathbb{E}\{|d|^2\}$ can be reformulated as
\begin{align}\label{eq: ap-c-17}\setcounter{equation}{33}	
&\mathbb{E}\{|d|^2\}=p_u^2\tau_p^2\mathbb{E}\left\{\left|\pmb{w}^H(\pmb{\Psi}_{mu}^{-1/2})^H\pmb{R}_{mu}^H\pmb{R}_{mu}\pmb{\Psi}_{mu}^{-1/2}\pmb{w}\right|^2\right\}\nonumber\\
	&\overset{(a)}{=}p_u^2\tau_p^2\tr(\pmb{\Gamma}_{mu})\tr(\pmb{\Gamma}_{mu})+p_u\tau_p\tr\left(\bar{\pmb{C}}_{mu}\pmb{\Gamma}_{mu}\right),
\end{align}
where $(a)$ invokes \eqref{eq: channel estimate matrix} and $\tr(\pmb{X}\pmb{Y})=\tr({\pmb{Y}\pmb{X}})$. Hence, by putting together all the above results, $\mathbb{E}\left\{\left|\hat{\pmb{h}}_{mu}^H\hat{\pmb{h}}_{mu}\right|^2\right\}$ can be expressed as
\begin{align}\label{eq: ap-c-18}
	&\bar{\pmb{h}}_{mu}^H\bar{\pmb{h}}_{mu}\bar{\pmb{h}}_{mu}^H\bar{\pmb{h}}_{mu}+3p_u\tau_p\bar{\pmb{h}}_{mu}^H\bar{\pmb{h}}_{mu}\tr\left(\pmb{\Gamma}_{mu}\right)\nonumber\\
	&+\tr(\bar{\pmb{h}}_{mu}^H\bar{\pmb{h}}_{mu}(\pmb{R}_{mu}^h-\pmb{\Lambda}_{mu}))+p_u^2\tau_p^2\tr(\pmb{\Gamma}_{mu})\tr(\pmb{\Gamma}_{mu})\nonumber\\
	&+p_u\tau_p\tr\left((\pmb{R}_{mu}^h-\pmb{\Lambda}_{mu})\pmb{\Gamma}_{mu}\right).\end{align}
By combining the above results associated with $m=n$ and $k=u$, we can arrive at
\begin{align}\label{eq: ap-c-19}
	\mathbb{E}\left\{\left|\hat{\pmb{h}}_{mu}^H\pmb{h}_{mu}\right|^2\right\}&=\left|\bar{\pmb{h}}_{mu}^H\bar{\pmb{h}}_{mu}\right|^2+3p_u\tau_p\bar{\pmb{h}}_{mu}^H\bar{\pmb{h}}_{mu}\tr\left(\pmb{\Gamma}_{mu}\right)\nonumber\\&+p_u^2\tau_p^2\left|\tr(\pmb{\Gamma}_{mu})\right|^2
	+\tr(\bar{\pmb{h}}_{mu}^H\bar{\pmb{h}}_{mu}\pmb{R}_{mu}^h)\nonumber\\
	&+p_u\tau_p\tr(\pmb{R}_{mu}^h\pmb{\Gamma}_{mu}).
\end{align}
Furthermore, we can compute the expectation term as $\tr\left(\mathbb{E}\left\{{\pmb{h}}_{mk}^H{\pmb{h}}_{mk}\hat{\pmb{h}}_{mu}^H\hat{\pmb{h}}_{mu}\right\}\right)=p_u\tau_p\bar{\pmb{h}}_{mk}^H\bar{\pmb{h}}_{mk}\tr(\pmb{\Gamma}_{mu})+\left|\bar{\pmb{h}}_{mu}^H\bar{\pmb{h}}_{mk}\right|^2+p_u\tau_p\tr(\pmb{R}_{mk}^h\pmb{\Gamma}_{mu})+\tr(\pmb{R}_{mk}^h)\bar{\pmb{h}}_{mu}^H\bar{\pmb{h}}_{mu}$, when $m=n$ and $k\notin\mathcal{P}_{u}$. In addition, for $m=n$ and $k\in\mathcal{P}_{u}\setminus\{u\}$, we can obtain $\mathbb{E}\left\{\left|\hat{\pmb{h}}_{mu}^H\pmb{h}_{mk}\right|^2\right\}=\mathbb{E}\left\{\left|\hat{\pmb{h}}_{mu}^H\hat{\pmb{h}}_{mk}\right|^2\right\}+\mathbb{E}\left\{\left|\hat{\pmb{h}}_{mu}^H\tilde{\pmb{h}}_{mk}\right|^2\right\}$ 
where we have $\mathbb{E}\left\{\left|\hat{\pmb{h}}_{mu}^H\tilde{\pmb{h}}_{mk}\right|^2\right\}=p_u\tau_p\tr\left(\pmb{\Gamma}_{mu}\pmb{\Lambda}_{mk}\right)+\tr\left(\bar{\pmb{h}}_{mu}^H\bar{\pmb{h}}_{mu}\pmb{\Lambda}_{mk}\right)$. {Similar to the calculation process of \eqref{eq: ap-c-12}, and bearing in mind that $\mathbb{E}\left\{\bar{\pmb{h}}_{mk}^H\pmb{R}_{mu}^h\pmb{\Psi}_{mu}^{-1}(\pmb{y}_{mu}^p-\bar{\pmb{y}}_{mu}^p)\right\}=0$, we can obtain the final result of $\mathbb{E}\left\{\left|\hat{\pmb{h}}_{mu}^H\hat{\pmb{h}}_{mk}\right|^2\right\}$ as in \eqref{eq: ap-c-26}, which is shown at the top of this page, where we have $a=\sqrt{p_u\tau_p}\pmb{w}^H\pmb{\Psi}_{mu}^{-1/2}\pmb{R}_{mu}^h\bar{\pmb{h}}_{mk}$, $b=\sqrt{p_u\tau_p}\bar{\pmb{h}}_{mu}^H\pmb{R}_{mk}^h\pmb{\Psi}_{mk}^{-1/2}\pmb{w}$ and $c=\pmb{w}^H\pmb{\Psi}_{mu}^{-1/2}\pmb{R}_{mu}\pmb{R}_{mk}\pmb{\Psi}_{mk}^{-1/2}\pmb{w}$ with $\pmb{w}\sim\mathcal{CN}(\pmb{0},\pmb{I}_J)$.} By combining the above results when $m=n$ and $k\in\mathcal{P}_{u}\setminus\{u\}$, we can derive
\begin{align}\label{eq: ap-c-24}\setcounter{equation}{37}	
\mathbb{E}&\left\{\left|\hat{\pmb{h}}_{mu}^H\pmb{h}_{mk}\right|^2\right\}=\left|\bar{\pmb{h}}_{mu}^H\bar{\pmb{h}}_{mk}\right|^2+p_u\tau_p\tr\left(\bar{\pmb{h}}_{mk}\bar{\pmb{h}}_{mk}^H\pmb{\Gamma}_{mu}\right)\nonumber\\
&+\tr\left(\bar{\pmb{h}}_{mu}\bar{\pmb{h}}_{mu}^H\pmb{R}^h_{mk}\right)
+p_u\tau_p\tr\left(\bar{\pmb{\Gamma}}_{muk}\right)\nonumber\\&+p_kp_u\tau_p^2\left|\tr\left(\pmb{R}_{mk}\pmb{\Psi}_{mu}^{-1}\pmb{R}_{mu}^h\right)\right|^2,
\end{align}
where $\bar{\pmb{\Gamma}}_{muk}=\pmb{\Gamma}_{mu}\pmb{R}_{mk}^h$.
By collecting all the above cases, $\mathbb{E}\left\{\left|\sum_{m=1}^M c_{mu}\hat{\pmb{h}}^H_{mu}{\pmb{h}}_{mk}\right|^2\right\}$ can be derived as \eqref{eq: ap-c-25}, which is provided at the top of this page. Finally, we substitute all the above-derived terms into \eqref{eq: SINR monte carlo}, and then \eqref{eq: SINR closed form} can be obtained.
\setcounter{eqnback}{\value{equation}} \setcounter{equation}{36}
\begin{figure*}[!t]
\begin{align}\label{eq: ap-c-26}
\mathbb{E}\left\{\left|\hat{\pmb{h}}_{mu}^H\hat{\pmb{h}}_{mk}\right|^2\right\}&=\left|\bar{\pmb{h}}_{mu}^H\bar{\pmb{h}}_{mk}\right|^2+\mathbb{E}\left\{\left|a\right|^2\right\}+\mathbb{E}\left\{\left|b\right|^2\right\}+p_u p_k\tau_p^2\mathbb{E}\left\{\left|c\right|^2\right\}\nonumber\\
&=\left|\bar{\pmb{h}}_{mu}^H\bar{\pmb{h}}_{mk}\right|^2+p_u\tau_p\tr\left(\bar{\pmb{h}}_{mk}\bar{\pmb{h}}_{mk}^H\pmb{\Gamma}_{mu}\right)+\tr\left(\bar{\pmb{h}}_{mu}\bar{\pmb{h}}_{mu}^H\bar{\pmb{C}}_{mk}\right)\nonumber\\
&+p_kp_u\tau_p^2\left|\tr\left(\pmb{R}_{mk}\pmb{\Psi}_{mu}^{-1}\pmb{R}_{mu}^h\right)\right|^2+p_u\tau_p\tr\left(\pmb{\Gamma}_{mu}\bar{\pmb{C}}_{mk}\right),
\end{align}
\hrulefill
\vspace{-2em}
\end{figure*}
\setcounter{eqncnt}{\value{equation}}
\setcounter{equation}{\value{eqnback}}
\setcounter{eqnback}{\value{equation}} \setcounter{equation}{38}
\begin{figure*}[!t]
\begin{align}\label{eq: ap-c-25}	
&\sum_{m=1}^M |c_{mu}|^2 \left[\left|\bar{\pmb{h}}_{mu}^H\bar{\pmb{h}}_{mk}\right|^2+p_u\tau_p\bar{\pmb{h}}_{mk}^H\pmb{\Gamma}_{mu}\bar{\pmb{h}}_{mk}+\bar{\pmb{h}}_{mu}^H\pmb{R}^h_{mk}\bar{\pmb{h}}_{mu}+p_u\tau_p\tr\left(\pmb{\Gamma}_{mu}\pmb{R}_{mk}^h\right)\right]\nonumber\\
	&+\begin{cases}
		\sum\limits_{m=1}^M\sum\limits_{n=1}^M c_{mu} c_{nu}^* p_u p_k\tau_p^2\tr(\pmb{R}_{mk}^h\pmb{\Psi}_{mu}^{-1}\pmb{R}_{mu}^h)\tr(\pmb{R}_{nk}^h\pmb{\Psi}_{nu}^{-1}\pmb{R}_{nu}^h),\ & k\in\mathcal{P}_u\setminus\{u\},\\
	\sum\limits_{m=1}^M\sum\limits_{n=1}^M c_{mu}c_{nu}^*\left[\tr\left(\bar{\pmb{h}}_{mu}^H\bar{\pmb{h}}_{mu}\right)\tr\left(\bar{\pmb{h}}_{nu}^H\bar{\pmb{h}}_{nu}\right)+p_u^2\tau_p^2\tr(\pmb{\Gamma}_{mu})\tr(\pmb{\Gamma}_{nu})\right.\\
	\left.+p_k\tau_p\tr\left(\bar{\pmb{h}}_{mu}^H\bar{\pmb{h}}_{mu}\right)\tr(\pmb{\Gamma}_{nu})+p_k\tau_p\tr\left(\bar{\pmb{h}}_{nu}^H\bar{\pmb{h}}_{nu}\right)\tr(\pmb{\Gamma}_{mu})\right],\ & k=u,\\
	0,\ & k\notin\mathcal{P}_u.
	\end{cases}\nonumber\\
	&+\begin{cases}
		\sum\limits_{m=1}^M|c_{mu}|^2p_kp_u\tau_p^2\left|\tr\left(\pmb{R}_{mk}\pmb{\Psi}_{mu}^{-1}\pmb{R}_{mu}^h\right)\right|^2,\ & k\in\mathcal{P}_u\setminus\{u\},\\	
		\sum\limits_{m=1}^M|c_{mu}|^2\left[2p_u\tau_p\bar{\pmb{h}}_{mu}^H\bar{\pmb{h}}_{mu}\tr\left(\pmb{\Gamma}_{mu}\right)+p_u^2\tau_p^2\left|\tr(\pmb{\Gamma}_{mu})\right|^2\right],\ & k=u,\\
		0,\ & k\notin\mathcal{P}_u.
		\end{cases}
\end{align}
\hrulefill
\vspace{-2em}
\end{figure*}
\setcounter{eqncnt}{\value{equation}}
\setcounter{equation}{\value{eqnback}}
\renewcommand{\refname}{References}
\mbox{} 
\nocite{*}
\bibliographystyle{IEEEtran}
\bibliography{paper_Zeping.bib}

\begin{thebibliography}{10}
\providecommand{\url}[1]{#1}
\csname url@samestyle\endcsname
\providecommand{\newblock}{\relax}
\providecommand{\bibinfo}[2]{#2}
\providecommand{\BIBentrySTDinterwordspacing}{\spaceskip=0pt\relax}
\providecommand{\BIBentryALTinterwordstretchfactor}{4}
\providecommand{\BIBentryALTinterwordspacing}{\spaceskip=\fontdimen2\font plus
\BIBentryALTinterwordstretchfactor\fontdimen3\font minus
  \fontdimen4\font\relax}
\providecommand{\BIBforeignlanguage}[2]{{%
\expandafter\ifx\csname l@#1\endcsname\relax
\typeout{** WARNING: IEEEtran.bst: No hyphenation pattern has been}%
\typeout{** loaded for the language `#1'. Using the pattern for}%
\typeout{** the default language instead.}%
\else
\language=\csname l@#1\endcsname
\fi
#2}}
\providecommand{\BIBdecl}{\relax}
\BIBdecl

\bibitem{7827017}
H.~Q. Ngo, A.~Ashikhmin, H.~Yang, E.~G. Larsson, and T.~L. Marzetta,
  ``Cell-free massive {MIMO} versus small cells,'' \emph{IEEE Trans. Wireless
  Commun.}, vol.~16, no.~3, pp. 1834--1850, Mar. 2017.

\bibitem{8097026}
H.~Q. Ngo, L.-N. Tran, T.~Q. Duong, M.~Matthaiou, and E.~G. Larsson, ``On the
  total energy efficiency of cell-free massive {MIMO},'' \emph{IEEE Trans.
  Green Commun. Netw}, vol.~2, no.~1, pp. 25--39, Mar. 2018.

\bibitem{zhang2020prospective}
J.~Zhang, E.~Bj{\"o}rnson, M.~Matthaiou, D.~W.~K. Ng, H.~Yang, and D.~J. Love,
  ``Prospective multiple antenna technologies for beyond {5G},'' \emph{IEEE J.
  Sel. Areas Commun.}, vol.~38, no.~8, pp. 1637--1660, Aug. 2020.

\bibitem{bjornson2019making}
E.~Bj{\"o}rnson and L.~Sanguinetti, ``Making cell-free massive {MIMO}
  competitive with {MMSE} processing and centralized implementation,''
  \emph{IEEE Trans. Wireless Commun.}, vol.~19, no.~1, pp. 77--90, Jan. 2019.

\bibitem{10522673}
H.~Q. Ngo, G.~Interdonato, E.~G. Larsson, G.~Caire, and J.~G. Andrews,
  ``Ultradense cell-free massive {MIMO for 6G}: Technical overview and open
  questions,'' \emph{Proceedings of the IEEE}, pp. 1--27, 2024.

\bibitem{6482234}
A.~Lozano, R.~W. Heath~Jr., and J.~G. Andrews, ``Fundamental limits of
  cooperation,'' \emph{IEEE Trans. Inf. Theory}, vol.~59, no.~9, pp.
  5213--5226, Sep. 2013.

\bibitem{9354156}
L.~Du, L.~Li, H.~Q. Ngo, T.~C. Mai, and M.~Matthaiou, ``Cell-free massive
  {MIMO}: Joint maximum-ratio and zero-forcing precoder with power control,''
  \emph{IEEE Trans. Commun.}, vol.~69, no.~6, pp. 3741--3756, Jun. 2021.

\bibitem{10197459}
Y.~Zhang, H.~Zhao, W.~Xia, W.~Xu, C.~Tang, and H.~Zhu, ``How much does
  reconfigurable intelligent surface improve cell-free massive {MIMO} uplink
  with hardware impairments?'' \emph{IEEE Trans. Commun.}, vol.~71, no.~11, pp.
  6677--6694, Nov. 2023.

\bibitem{9614196}
J.~An, C.~Xu, L.~Gan, and L.~Hanzo, ``Low-complexity channel estimation and
  passive beamforming for {RIS}-assisted {MIMO} systems relying on discrete
  phase shifts,'' \emph{IEEE Trans. Commun.}, vol.~70, no.~2, pp. 1245--1260,
  Feb. 2022.

\bibitem{li2022reconfigurable}
Q.~Li, M.~El-Hajjar, I.~Hemadeh, A.~Shojaeifard, A.~A. Mourad, B.~Clerckx, and
  L.~Hanzo, ``Reconfigurable intelligent surfaces relying on non-diagonal phase
  shift matrices,'' \emph{IEEE Trans. Veh. Technol.}, vol.~71, no.~6, pp.
  6367--6383, Jun. 2022.

\bibitem{8910627}
Q.~Wu and R.~Zhang, ``Towards smart and reconfigurable environment: Intelligent
  reflecting surface aided wireless network,'' \emph{{IEEE} Commun. Mag.},
  vol.~58, no.~1, pp. 106--112, Jan. 2020.

\bibitem{huang2019reconfigurable}
C.~Huang, A.~Zappone, G.~C. Alexandropoulos, M.~Debbah, and C.~Yuen,
  ``Reconfigurable intelligent surfaces for energy efficiency in wireless
  communication,'' \emph{IEEE Trans. Wireless Commun.}, vol.~18, no.~8, pp.
  4157--4170, Aug. 2019.

\bibitem{van2021reconfigurable}
T.~Van~Chien, H.~Q. Ngo, S.~Chatzinotas, M.~Di~Renzo, and B.~Ottersten,
  ``Reconfigurable intelligent surface-assisted cell-free massive {MIMO}
  systems over spatially-correlated channels,'' \emph{IEEE Trans. Wireless
  Commun.}, vol.~21, no.~7, pp. 5106--5128, Jul. 2021.

\bibitem{10167480}
E.~Shi, J.~Zhang, D.~W.~K. Ng, and B.~Ai, ``Uplink performance of {RIS}-aided
  cell-free massive {MIMO} system with electromagnetic interference,''
  \emph{IEEE J. Sel. Areas Commun.}, vol.~41, no.~8, pp. 2431--2445, Aug. 2023.

\bibitem{10326460}
J.~Dai, J.~Ge, K.~Zhi, C.~Pan, Z.~Zhang, J.~Wang, and X.~You, ``Two-timescale
  transmission design for {RIS}-aided cell-free massive {MIMO} systems,''
  \emph{IEEE Trans. Wireless Commun.}, pp. 1--20, Nov. 2023.

\bibitem{10129196}
J.~Yao, J.~Xu, W.~Xu, D.~W.~K. Ng, C.~Yuen, and X.~You, ``Robust beamforming
  design for {RIS}-aided cell-free systems with {CSI} uncertainties and
  capacity-limited backhaul,'' \emph{IEEE Trans. Commun.}, vol.~71, no.~8, pp.
  4636--4649, Aug. 2023.

\bibitem{shi2022wireless}
E.~Shi, J.~Zhang, S.~Chen, J.~Zheng, Y.~Zhang, D.~W.~K. Ng, and B.~Ai,
  ``Wireless energy transfer in {RIS}-aided cell-free massive {MIMO} systems:
  Opportunities and challenges,'' \emph{IEEE Commun. Mag.}, vol.~60, no.~3, pp.
  26--32, Mar. 2022.

\bibitem{10468556}
E.~Shi, J.~Zhang, J.~Zheng, B.~Ai, and D.~W.~K. Ng, ``{RIS}-aided cell-free
  massive {MIMO} systems with channel aging,'' \emph{IEEE Trans. Veh.
  Technol.}, pp. 1--16, 2024.

\bibitem{9685730}
D.~L. Galappaththige, D.~Kudathanthirige, and G.~Amarasuriya, ``Performance
  analysis of {IRS}-assisted cell-free communication,'' in \emph{2021 IEEE
  Global Communications Conference ({GLOBECOM})}, 2021, pp. 1--6.

\bibitem{9473521}
B.~Al-Nahhas, M.~Obeed, A.~Chaaban, and M.~J. Hossain, ``{RIS}-aided cell-free
  massive {MIMO}: Performance analysis and competitiveness,'' in \emph{2021
  IEEE International Conference on Communications Workshops ({ICC} Workshops)},
  2021, pp. 1--6.

\bibitem{9459505}
Z.~Zhang and L.~Dai, ``A joint precoding framework for wideband reconfigurable
  intelligent surface-aided cell-free network,'' \emph{IEEE Trans. Signal
  Process.}, vol.~69, pp. 4085--4101, Jun. 2021.

\bibitem{10290997}
B.~Li, Y.~Hu, Z.~Dong, E.~Panayirci, H.~Jiang, and Q.~Wu, ``Energy-efficient
  design for reconfigurable intelligent surface aided cell-free ultra dense
  hetnets,'' \emph{IEEE Trans. Veh. Technol.}, vol.~73, no.~3, pp. 3767--3785,
  Mar. 2024.

\bibitem{10175060}
S.-N. Jin, D.-W. Yue, Y.-L. Chen, and Q.~Hu, ``Energy efficiency maximization
  in {IRS}-aided cell-free massive {MIMO} system,'' \emph{IEEE Wireless Commun.
  Lett.}, vol.~12, no.~10, pp. 1652--1656, Oct. 2023.

\bibitem{9363171}
Q.~N. Le, V.-D. Nguyen, O.~A. Dobre, and R.~Zhao, ``Energy efficiency
  maximization in {RIS}-aided cell-free network with limited backhaul,''
  \emph{IEEE Commun. Lett.}, vol.~25, no.~6, pp. 1974--1978, Jun. 2021.

\bibitem{10035459}
W.~Lyu, Y.~Xiu, S.~Yang, C.~Yuen, and Z.~Zhang, ``Energy-efficient cell-free
  network assisted by hybrid {RISs},'' \emph{IEEE Wireless Commun. Lett.},
  vol.~12, no.~4, pp. 718--722, Apr. 2023.

\bibitem{9499059}
R.~Karasik, O.~Simeone, M.~Di~Renzo, and S.~Shamai~(Shitz), ``Adaptive coding
  and channel shaping through reconfigurable intelligent surfaces: An
  information-theoretic analysis,'' \emph{IEEE Trans. Commun.}, vol.~69,
  no.~11, pp. 7320--7334, Nov. 2021.

\bibitem{8941126}
W.~Yan, X.~Yuan, and X.~Kuai, ``Passive beamforming and information transfer
  via large intelligent surface,'' \emph{IEEE Wireless Commun. Lett.}, vol.~9,
  no.~4, pp. 533--537, Apr. 2020.

\bibitem{9133134}
S.~Guo, S.~Lv, H.~Zhang, J.~Ye, and P.~Zhang, ``Reflecting modulation,''
  \emph{IEEE J. Sel. Areas Commun.}, vol.~38, no.~11, pp. 2548--2561, Nov.
  2020.

\bibitem{9217944}
S.~Lin, B.~Zheng, G.~C. Alexandropoulos, M.~Wen, M.~D. Renzo, and F.~Chen,
  ``Reconfigurable intelligent surfaces with reflection pattern modulation:
  Beamforming design and performance analysis,'' \emph{IEEE Trans. Wireless
  Commun.}, vol.~20, no.~2, pp. 741--754, Feb. 2021.

\bibitem{9516949}
S.~Lin, F.~Chen, M.~Wen, Y.~Feng, and M.~Di~Renzo, ``Reconfigurable intelligent
  surface-aided quadrature reflection modulation for simultaneous passive
  beamforming and information transfer,'' \emph{IEEE Trans. Wireless Commun.},
  vol.~21, no.~3, pp. 1469--1481, Mar. 2022.

\bibitem{10224829}
J.~Yao, J.~Xu, W.~Xu, C.~Yuen, and X.~You, ``Superimposed {RIS}-phase
  modulation for {MIMO} communications: A novel paradigm of information
  transfer,'' \emph{IEEE Trans. Wireless Commun.}, vol.~23, no.~4, pp.
  2978--2993, Apr. 2024.

\bibitem{9965423}
Z.~Yigit, E.~Basar, M.~Wen, and I.~Altunbas, ``Hybrid reflection modulation,''
  \emph{IEEE Trans. Wireless Commun.}, vol.~22, no.~6, pp. 4106--4116, Jun.
  2023.

\bibitem{9110889}
B.~Di, H.~Zhang, L.~Song, Y.~Li, Z.~Han, and H.~V. Poor, ``Hybrid beamforming
  for reconfigurable intelligent surface based multi-user communications:
  Achievable rates with limited discrete phase shifts,'' \emph{IEEE J. Sel.
  Areas Commun.}, vol.~38, no.~8, pp. 1809--1822, Aug. 2020.

\bibitem{9779130}
E.~Shi, J.~Zhang, R.~He, H.~Jiao, Z.~Wang, B.~Ai, and D.~W.~K. Ng, ``Spatially
  correlated reconfigurable intelligent surfaces-aided cell-free massive {MIMO}
  systems,'' \emph{IEEE Trans. Veh. Technol.}, vol.~71, no.~8, pp. 9073--9077,
  Aug. 2022.

\bibitem{9507331}
Z.~Sui, S.~Yan, H.~Zhang, L.-L. Yang, and L.~Hanzo, ``Approximate message
  passing algorithms for low complexity {OFDM-IM} detection,'' \emph{IEEE
  Trans. Veh. Technol.}, vol.~70, no.~9, pp. 9607--9612, Sep. 2021.

\bibitem{10183832}
Z.~Sui, S.~Yan, H.~Zhang, S.~Sun, Y.~Zeng, L.-L. Yang, and L.~Hanzo,
  ``Performance analysis and approximate message passing detection of
  orthogonal time sequency multiplexing modulation,'' \emph{IEEE Trans.
  Wireless Commun.}, vol.~23, no.~3, pp. 1913--1928, Mar. 2024.

\bibitem{hoydis2013massive}
J.~Hoydis, S.~Ten~Brink, and M.~Debbah, ``Massive {MIMO} in the {UL/DL} of
  cellular networks: How many antennas do we need?'' \emph{IEEE J. Sel. Areas
  Commun.}, vol.~31, no.~2, pp. 160--171, Feb. 2013.

\bibitem{bjornson2020rayleigh}
E.~Bj{\"o}rnson and L.~Sanguinetti, ``Rayleigh fading modeling and channel
  hardening for reconfigurable intelligent surfaces,'' \emph{IEEE Wireless
  Commun. Lett.}, vol.~10, no.~4, pp. 830--834, Apr. 2020.

\bibitem{9973349}
K.~Zhi, C.~Pan, H.~Ren, K.~Wang, M.~Elkashlan, M.~D. Renzo, R.~Schober, H.~V.
  Poor, J.~Wang, and L.~Hanzo, ``Two-timescale design for reconfigurable
  intelligent surface-aided massive {MIMO} systems with imperfect {CSI},''
  \emph{{IEEE} Trans. Inform. Theory}, vol.~69, no.~5, pp. 3001--3033, 2023.

\bibitem{ozdogan2019performance}
{\"O}.~{\"O}zdogan, E.~Bj{\"o}rnson, and J.~Zhang, ``Performance of cell-free
  massive {MIMO} with {Rician} fading and phase shifts,'' \emph{IEEE Trans.
  Wireless Commun.}, vol.~18, no.~11, pp. 5299--5315, Nov. 2019.

\bibitem{cover1999elements}
T.~M. Cover, \emph{Elements of Information Theory}.\hskip 1em plus 0.5em minus
  0.4em\relax John Wiley \& Sons, 1999.

\bibitem{femenias2021swipt}
G.~Femenias, J.~Garc{\'\i}a-Morales, and F.~Riera-Palou, ``{SWIPT}-enhanced
  cell-free massive {MIMO} networks,'' \emph{IEEE Trans. Commun.}, vol.~69,
  no.~8, pp. 5593--5607, Aug. 2021.

\bibitem{10480438}
J.~Wang, W.~Tang, J.~C. Liang, L.~Zhang, J.~Y. Dai, X.~Li, S.~Jin, Q.~Cheng,
  and T.~J. Cui, ``Reconfigurable intelligent surface: Power consumption
  modeling and practical measurement validation,'' \emph{IEEE Trans. Commun.},
  pp. 1--13, 2024.

\bibitem{caponetto2003chaotic}
R.~Caponetto, L.~Fortuna, S.~Fazzino, and M.~G. Xibilia, ``Chaotic sequences to
  improve the performance of evolutionary algorithms,'' \emph{IEEE Trans. Evol.
  Comput.}, vol.~7, no.~3, pp. 289--304, Jun. 2003.

\bibitem{985692}
M.~Clerc and J.~Kennedy, ``The particle swarm: Explosion, stability, and
  convergence in a multidimensional complex space,'' \emph{IEEE Trans. Evol.
  Comput.}, vol.~6, no.~1, pp. 58--73, Feb. 2002.

\bibitem{9737367}
Z.~Wang, J.~Zhang, B.~Ai, C.~Yuen, and M.~Debbah, ``Uplink performance of
  cell-free massive {MIMO} with multi-antenna users over jointly-correlated
  rayleigh fading channels,'' \emph{IEEE Trans. Wireless Commun.}, vol.~21,
  no.~9, pp. 7391--7406, Sep. 2022.

\bibitem{zhang2015multivariate}
J.~Zhang, M.~Matthaiou, G.~K. Karagiannidis, and L.~Dai, ``On the multivariate
  gamma--gamma distribution with arbitrary correlation and applications in
  wireless communications,'' \emph{IEEE Trans. Veh. Technol.}, vol.~65, no.~5,
  pp. 3834--3840, May 2015.

\end{thebibliography}
\begin{IEEEbiography}[{\includegraphics[width=1in,height=1.25in,clip,keepaspectratio]{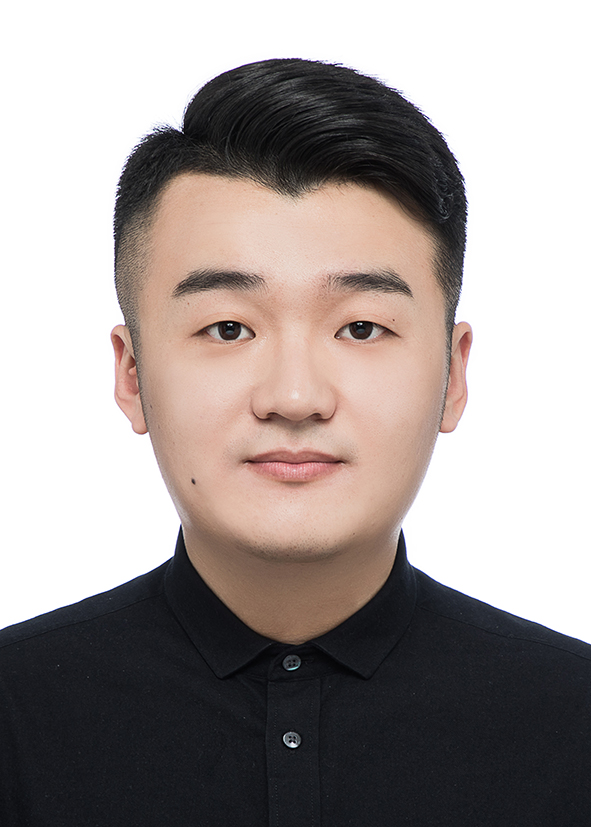}}]{Zeping Sui}
(Member, IEEE) received the B. Eng. degree in electronic information engineering from Northwestern Polytechnical University, Xi' an, China, in 2017, and the Ph.D. degree in signal and information processing from the University of Chinese Academy of Sciences, Beijing, China, in 2023. From March 2022 to Feb 2023, He was a research assistant with the Institute for Infocomm Research (I2R), Agency for Science, Technology and Research (A*STAR), Singapore. He is currently a research fellow with the Centre for Wireless Innovation (CWI), Queen's University Belfast, Belfast, UK. His research interests include 6G wireless communication networks, statistical signal processing, system performance analysis and optimization, and compressed sensing.
\end{IEEEbiography}
\begin{IEEEbiography}[{\includegraphics[width=1in,height=1.25in,clip,keepaspectratio]{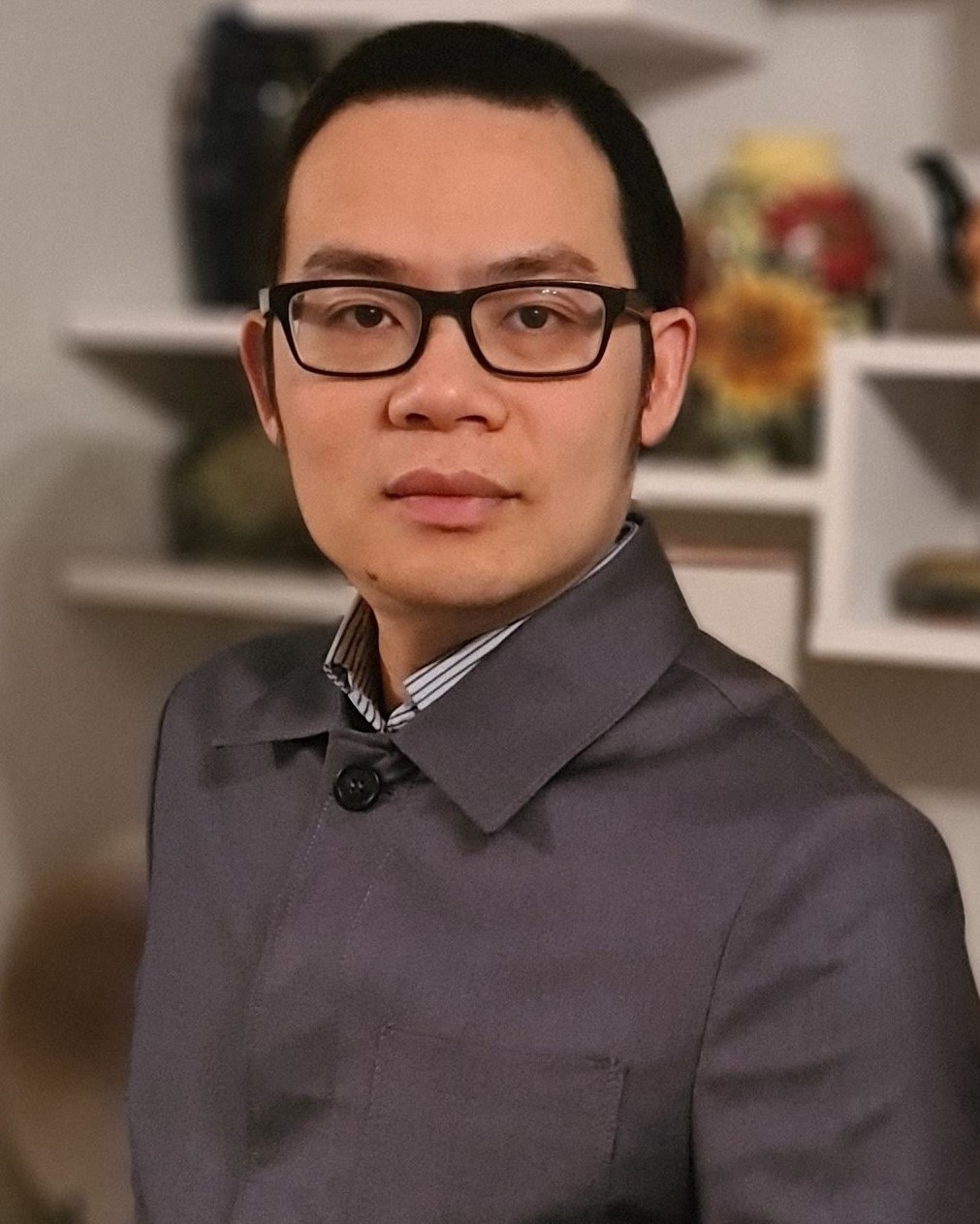}}] {Hien Quoc Ngo} (Senior Member, IEEE) received the B.S. degree in electrical engineering from the Ho Chi Minh City University of Technology, Vietnam, in 2007, the M.S. degree in electronics and radio engineering from Kyung Hee University, South Korea, in 2010, and the Ph.D. degree in communication systems from Link\"oping University (LiU), Sweden, in 2015. In 2014, he visited the Nokia Bell Labs, Murray Hill, New Jersey, USA. From January 2016 to April 2017, Hien Quoc Ngo was a VR researcher at the Department of Electrical Engineering (ISY), LiU. He was also a Visiting Research Fellow at the School of Electronics, Electrical Engineering and Computer Science, Queen's University Belfast, UK, funded by the Swedish Research Council.

Hien Quoc Ngo is currently a Reader (Associate Professor) at Queen's University Belfast, UK. His main research interests include massive MIMO systems, cell-free massive MIMO, reconfigurable intelligent surfaces, physical layer security, and cooperative communications. He has co-authored many research papers in wireless communications and co-authored the Cambridge University Press textbook \emph{Fundamentals of Massive MIMO} (2016).

Dr. Hien Quoc Ngo received the IEEE ComSoc Stephen O. Rice Prize in 2015, the IEEE ComSoc Leonard G. Abraham Prize in 2017,  the Best Ph.D. Award from EURASIP in 2018, and the IEEE CTTC Early Achievement Award in 2023. He also received the IEEE Sweden VT-COM-IT Joint Chapter Best Student Journal Paper Award in 2015. He was an \emph{IEEE Communications Letters} exemplary reviewer for 2014, an \emph{IEEE Transactions on Communications} exemplary reviewer for 2015, and an \emph{IEEE Wireless Communications Letters} exemplary reviewer for 2016.  He was awarded the UKRI Future Leaders Fellowship in 2019.
Dr. Hien Quoc Ngo currently serves as an Editor for the IEEE Transactions on Wireless Communications, the IEEE Wireless Communications Letters, Digital Signal Processing, Elsevier Physical Communication (PHYCOM). He was a Guest Editor of IET Communications, special issue on ``Recent Advances on 5G Communications'' and a Guest Editor of  IEEE Access, special issue on ``Modelling, Analysis, and Design of 5G Ultra-Dense Networks'', in 2017. He has been a member of Technical Program Committees for many IEEE conferences such as ICC, GLOBECOM, WCNC, and VTC.
\end{IEEEbiography}
\begin{IEEEbiography}[{\includegraphics[width=1in,height=1.25in,clip,keepaspectratio]{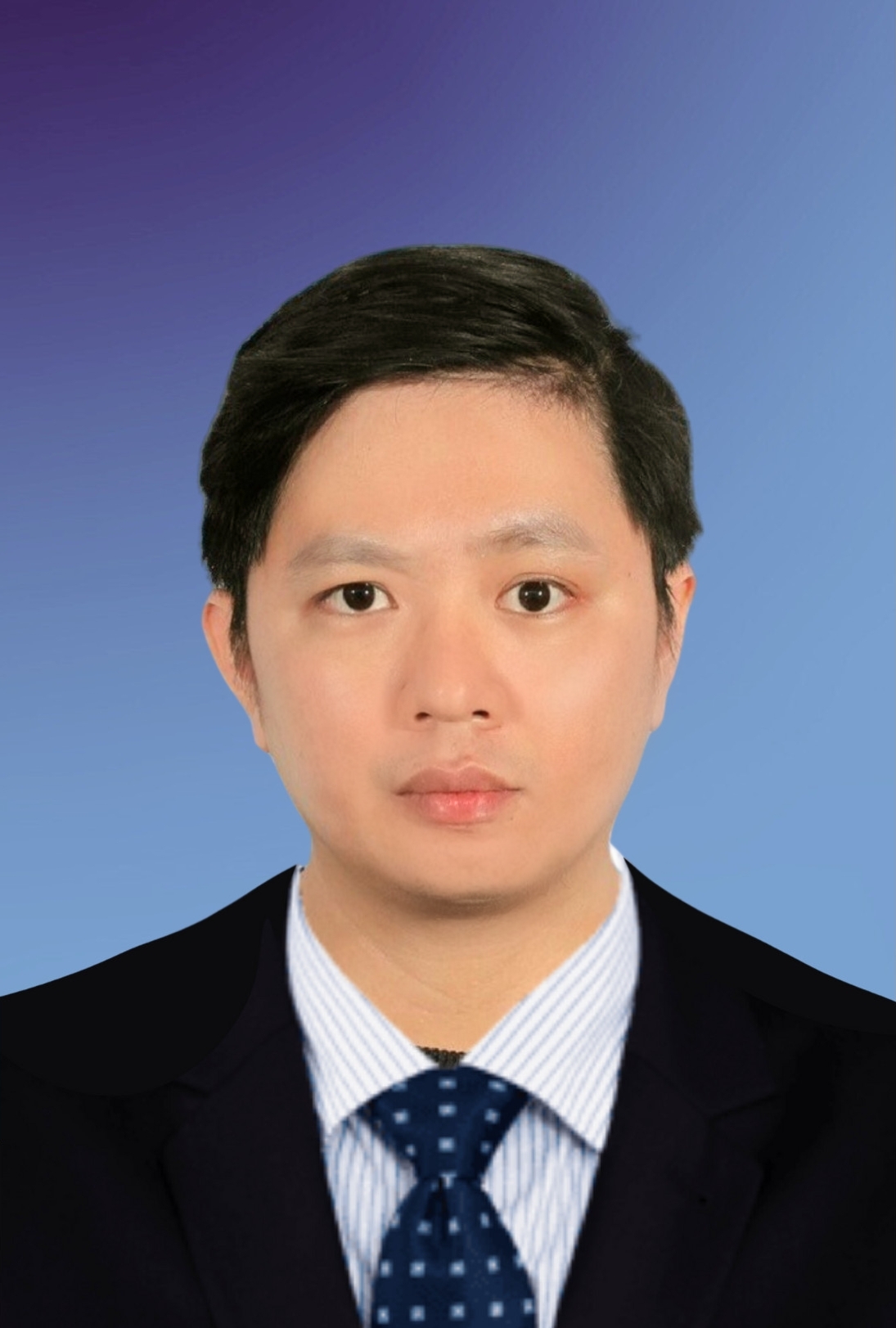}}] {Trinh Van Chien} received the B.S. degree in electronics and telecommunications from the Hanoi University of Science and Technology (HUST), Hanoi, Vietnam, in 2012, the M.S. degree in electrical and computer engineering from Sungkyunkwan University (SKKU), Seoul, South Korea, in 2014, and the Ph.D. degree in communication systems from Link\"oping University (LiU), Link\"oping, Sweden, in 2020. He was a Research Associate with the University of Luxembourg, Esch-sur-Alzette, Luxembourg. He is currently with the School of Information and Communication Technology (SoICT), HUST. His interest lies in convex optimization problems and machine learning applications for wireless communications and image \& video processing. Dr. Chien received the Award of Scientific Excellence in the first year of the 5G wireless project funded by European Union Horizon 2020. He was an  Exemplary Reviewer of IEEE Wireless Communications Letters Exemplary Reviewer in 2016, 2017, and 2021, and IEEE Transactions on Communications Exemplary Reviewer in 2022. He won the Science \& Technology Award in 2023. 
\end{IEEEbiography}

\begin{IEEEbiography}[{\includegraphics[width=1in,height=1.25in,clip,keepaspectratio]{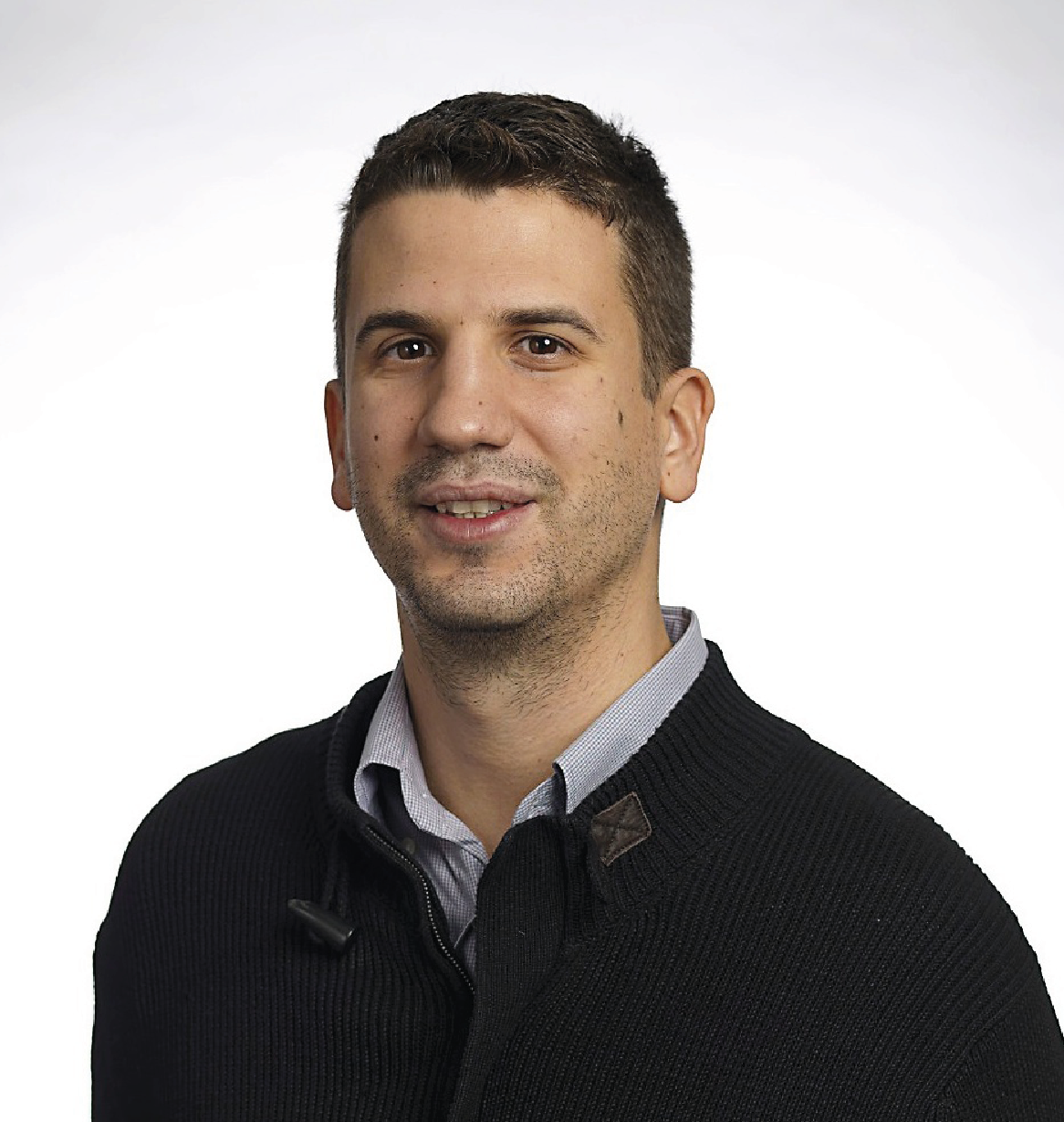}}]
{Michail Matthaiou}(Fellow, IEEE) was born in Thessaloniki, Greece in 1981. He obtained the Diploma degree (5 years) in Electrical and Computer Engineering from the Aristotle University of Thessaloniki, Greece in 2004. He then received the M.Sc. (with distinction) in Communication Systems and Signal Processing from the University of Bristol, U.K. and Ph.D. degrees from the University of Edinburgh, U.K. in 2005 and 2008, respectively. From September 2008 through May 2010, he was with the Institute for Circuit Theory and Signal Processing, Munich University of Technology (TUM), Germany working as a Postdoctoral Research Associate. He is currently a Professor of Communications Engineering and Signal Processing and Deputy Director of the Centre for Wireless Innovation (CWI) at Queen's University Belfast, U.K. after holding an Assistant Professor position at Chalmers University of Technology, Sweden. His research interests span signal processing for wireless communications, beyond massive MIMO, intelligent reflecting surfaces, mm-wave/THz systems and deep learning for communications.

Dr. Matthaiou and his coauthors received the IEEE Communications Society (ComSoc) Leonard G. Abraham Prize in 2017. He currently holds the ERC Consolidator Grant BEATRICE (2021-2026) focused on the interface between information and electromagnetic theories. To date, he has received the prestigious 2023 Argo Network Innovation Award, the 2019 EURASIP Early Career Award and the 2018/2019 Royal Academy of Engineering/The Leverhulme Trust Senior Research Fellowship. His team was also the Grand Winner of the 2019 Mobile World Congress Challenge. He was the recipient of the 2011 IEEE ComSoc Best Young Researcher Award for the Europe, Middle East and Africa Region and a co-recipient of the 2006 IEEE Communications Chapter Project Prize for the best M.Sc. dissertation in the area of communications. He has co-authored papers that received best paper awards at the 2018 IEEE WCSP and 2014 IEEE ICC. In 2014, he received the Research Fund for International Young Scientists from the National Natural Science Foundation of China. He is currently the Editor-in-Chief of Elsevier Physical Communication, a Senior Editor for \textsc{IEEE Wireless Communications Letters} and \textsc{IEEE Signal Processing Magazine}, and an Associate Editor for \textsc{IEEE Transactions on Communications}. He is an IEEE and AAIA Fellow.
\end{IEEEbiography}
\begin{IEEEbiography}[{\includegraphics[width=1in,height=1.25in,clip,keepaspectratio]{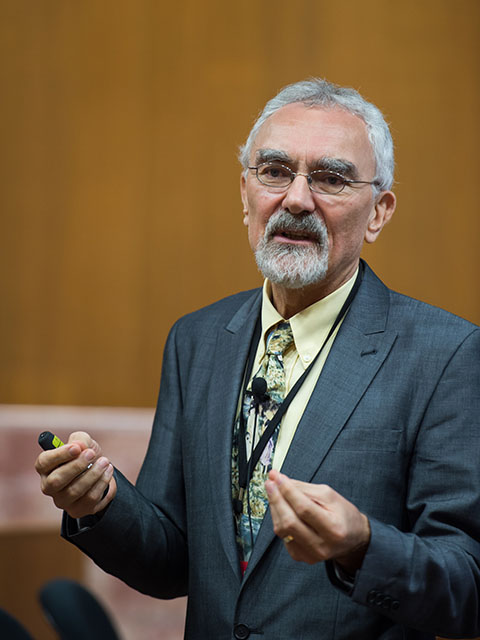}}]
{Lajos Hanzo}(Life Fellow, IEEE) received Honorary Doctorates  from the Technical University of Budapest (2009) and Edinburgh University (2015). He is a Foreign Member of the Hungarian Science-Academy, Fellow of the Royal Academy of Engineering (FREng), of the IET, of EURASIP and holds the IEEE Eric Sumner Technical Field Award. For further details please see \href{http://www-mobile.ecs.soton.ac.uk/}{http://www-mobile.ecs.soton.ac.uk/}, \href{http://en.wikipedia.org/wiki/Lajos\_Hanzo}{http://en.wikipedia.org/wiki/Lajos\_Hanzo}\end{IEEEbiography}

\end{document}